\documentclass[letterpaper,11pt]{article}

\usepackage[letterpaper,margin=1in]{geometry}

\usepackage{mathptmx}

\usepackage{tcolorbox}
\usepackage{fancyhdr}
\usepackage{float}
\usepackage{pdfpages}
\usepackage{relsize}
\usepackage{graphicx}

\newenvironment{titemize}{
\begin{itemize}
\setlength{\itemsep}{1pt}
\setlength{\parskip}{0pt}
\setlength{\parsep}{0pt}
}
{
\end{itemize}
}

\usepackage[hyphens]{url}
\usepackage{wrapfig}
\usepackage{hyperref}
\usepackage{graphicx}
\usepackage[numbers,sort,square]{natbib}
\usepackage{doi}
\usepackage{color}
\definecolor{linkblue}{RGB}{0,0,180}
\hypersetup{
     colorlinks=true,
     citecolor=linkblue,
     filecolor=black,
     linkcolor=linkblue,
     urlcolor=linkblue
}

\usepackage{datetime}

\usepackage{titlesec}
\titlespacing{\paragraph} {0pt}{7pt}{5pt}

\begin{document}

\pagestyle{empty}
\includepdf[pages=1]{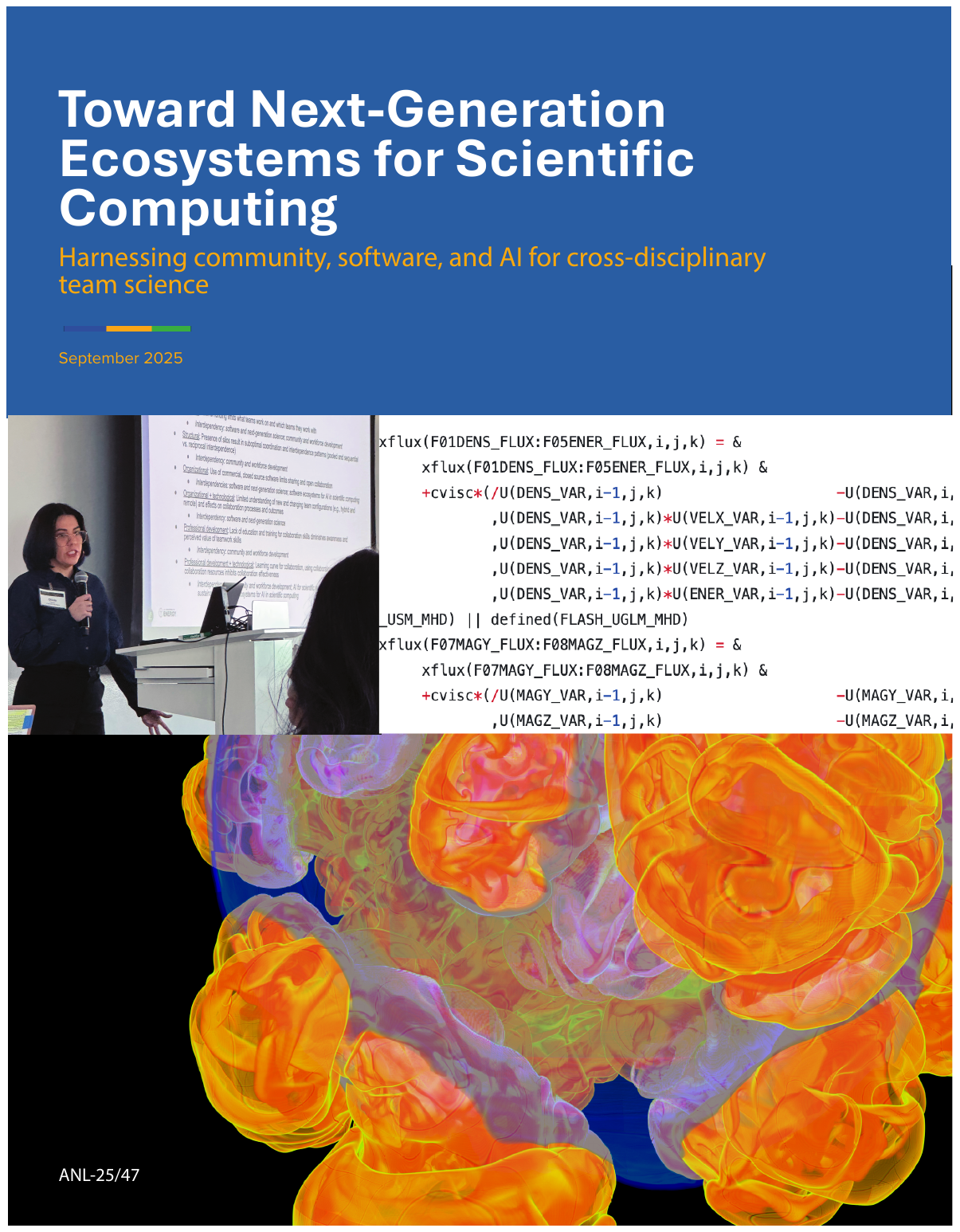}
\includepdf[pages=1]{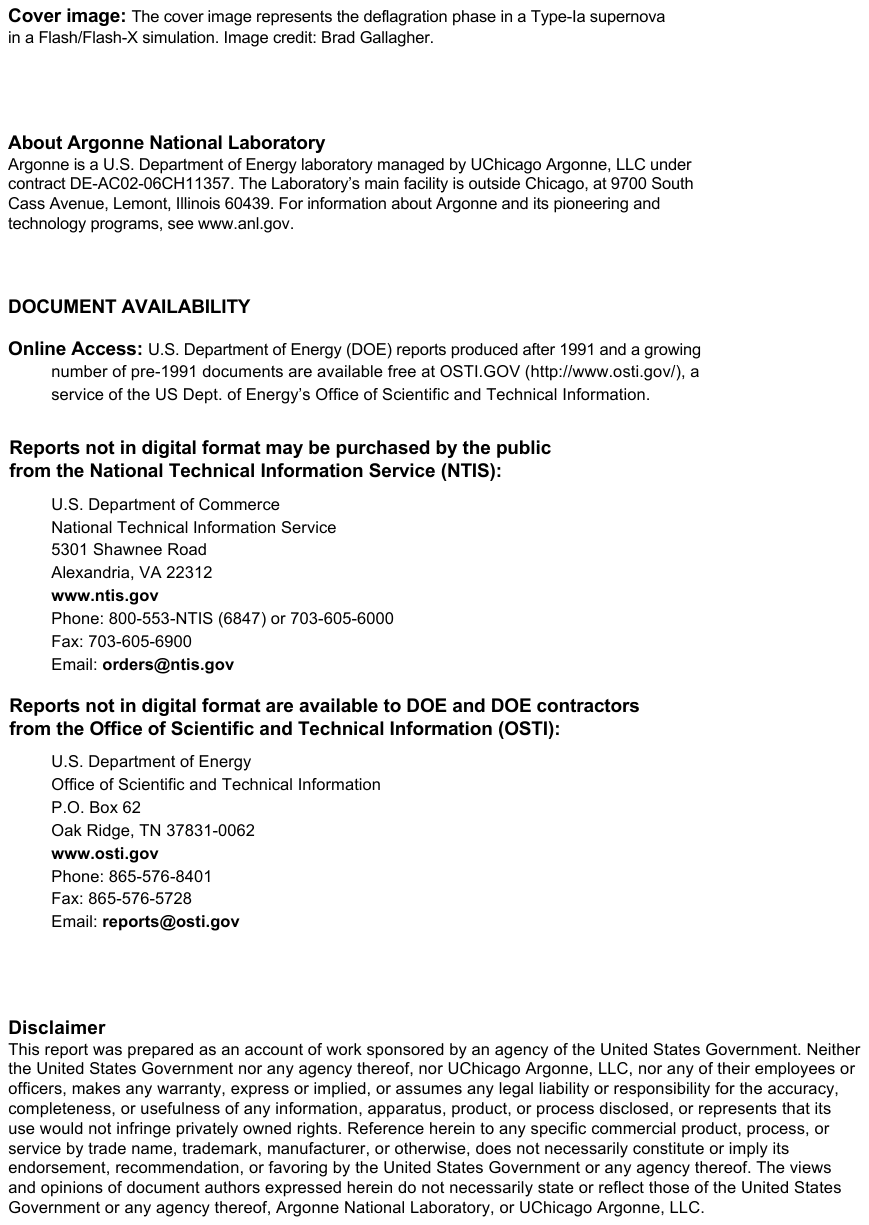}
\clearpage

\begin{center}
$\,$
\vspace{1.0in}

{\larger[2] \bf Report of the 2025 Workshop on}

\bigskip

{\larger[3] \bf Next-Generation Ecosystems for Scientific Computing:} \\

\smallskip

{\larger[1] \bf Harnessing Community, Software, and AI for Cross-Disciplinary Team Science}
\end{center}

\begin{center}

{\larger[1]

{\bf September 26, 2025}

\bigskip
\bigskip
{\bf Workshop Location:}\\
Chicago, IL

\bigskip
\bigskip

{\bf Workshop Dates:}\\
April 29 -- May 1, 2025
\bigskip
\bigskip
\bigskip
\bigskip
}
\end{center}
{\bf Suggested Citation}: Report of the 2025 Workshop on Next-Generation Ecosystems for Scientific Computing: Harnessing Community, Software, and AI for Cross-Disciplinary Team Science. L.C. McInnes, D.~Arnold, P.~Balaprakash, M. Bernhardt, B. Cerny, A. Dubey, R. Giles, D.W. Hood, M.A. Leung, V. L\'opez-Marrero, P.~Messina, O.B. Newton, C. Oehmen, S.M. Wild, J. Willenbring, L. Woodley, T. Baylis, D.E. Bernholdt, C. Camaño, J. Cohoon, C. Ferenbaugh, S.M. Fiore, S. Gesing, D. Gómez-Zará, J. Howison, T. Islam, D. Kepczynski, C. Lively, H. Menon, B. Messer, M. Ngom, U. Paliath, M.E. Papka, I. Qualters, E.M. Raybourn, K. Riley, P. Rodriguez, D. Rouson, M. Schwalbe, S.K. Seal, \"{O}. S\"{u}rer, V. Taylor, and L. Wu.
Report ANL-25/47, 2025, 
\url{https://doi.org/10.48550/arXiv.2510.03413}.

\bigskip
\bigskip
\bigskip
\begin{center}
\vspace{-0.1in}
\includegraphics[width=1.0\textwidth] {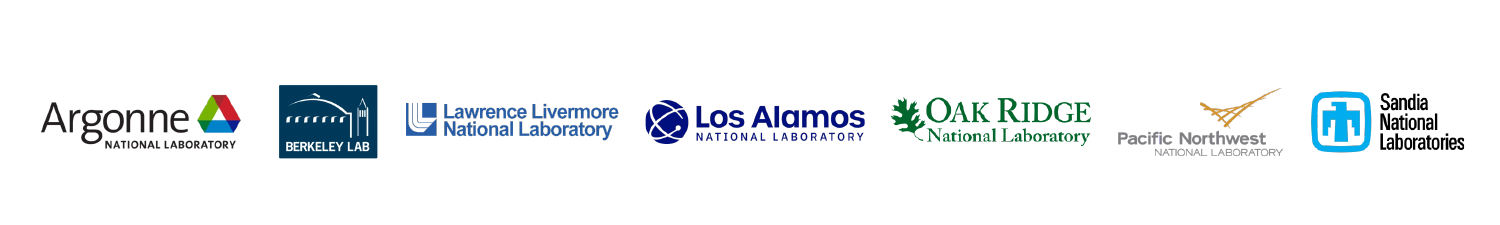}
\end{center}

\newpage

{\large
\begin{center}

{\bf Report Authors:}

\bigskip

Lois Curfman McInnes, Argonne National Laboratory\\
Dorian Arnold, Emory University\\
Prasanna Balaprakash, Oak Ridge National Laboratory\\
Mike Bernhardt, Team Libra\\
Beth Cerny, Argonne National Laboratory\\
Anshu Dubey, Argonne National Laboratory\\
Roscoe Giles, Boston University\\
Denice Ward Hood, University of Illinois Urbana-Champaign\\
Mary Ann Leung, Sustainable Horizons Institute\\
Vanessa L\'opez-Marrero, Stony Brook University\\
Paul Messina, Argonne National Laboratory (retired)\\
Olivia B.\ Newton, University of Montana\\
Chris Oehmen, Pacific Northwest National Laboratory\\
Stefan M.\ Wild, Lawrence Berkeley National Laboratory\\
Jim Willenbring, Sandia National Laboratories\\
Lou Woodley, Center for Scientific Collaboration and Community Engagement\\
Tony Baylis, Lawrence Livermore National Laboratory\\
David E. Bernholdt, Oak Ridge National Laboratory\\
Chris Cama\~{n}o, California Institute of Technology\\
Johannah Cohoon, Lawrence Berkeley National Laboratory\\
Charles Ferenbaugh, Los Alamos National Laboratory\\
Stephen M. Fiore, University of Central Florida\\
Sandra Gesing, US Research Software Engineers Association\\
Diego G\'{o}mez-Zar\'{a}, University of Notre Dame\\
James Howison, University of Texas at Austin\\
Tanzima Islam, Texas State University\\
David Kepczynski, Ford Motor Company\\
Charles Lively, Lawrence Berkeley National Laboratory\\
Harshitha Menon, 
Lawrence Livermore National Laboratory\\
Bronson Messer,  Oak Ridge National Laboratory\\
Marieme Ngom, Argonne National Laboratory\\
Umesh Paliath, GE Aerospace Research\\
Michael E. Papka, Argonne National Laboratory,  University of Illinois Chicago\\
Irene Qualters, Los Alamos National Laboratory (retired)\\
Elaine M. Raybourn, Sandia National Laboratories\\
Katherine Riley, Argonne National Laboratory\\
Paulina Rodriguez, The George Washington University\\
Damian Rouson, Lawrence Berkeley National Laboratory\\
Michelle Schwalbe, National Academies of Sciences\\
Sudip K. Seal, Oak Ridge National Laboratory\\
\"{O}zge S\"{u}rer, Miami University\\
Valerie Taylor, Argonne National Laboratory\\
Lingfei Wu, University of Pittsburgh\\

\end{center}
}
\clearpage

\pagestyle{plain}
\pagenumbering{roman}
\newpage
\tableofcontents

\newpage
\phantomsection
\section*{Executive Summary: Reimagining Scientific Computing in the Age of AI}

\addcontentsline{toc}{section}{Executive Summary: Reimagining Scientific Computing in the Age of AI}



The 2025 {\em Workshop on Next-Generation Ecosystems for Scientific Computing: Harnessing Community, Software, and AI for Cross-Disciplinary Team Science} convened  in Chicago, IL, from April 29 to May 1, 2025. 
The event brought together more than 40 experts from high-performance computing (HPC), AI, computational science, software engineering, applied mathematics, social sciences, and 
community development to chart a path toward more powerful, sustainable, and collaborative scientific software ecosystems.

Workshop discussions underscored a powerful truth: scientific computing is at a turning point. As AI grows more capable and scientific questions become more complex, a transformation is unavoidable. The traditional ways we compute, code, and collaborate are no longer adequate.
Periods of profound transformation bring both unprecedented opportunity and inherent uncertainty. Hence, addressing the complexity of this moment requires more than incremental progress—it necessitates strategic risk-taking and foundational shifts in approach. Just as the rise of digital computing reshaped nearly every scientific and industrial domain, AI-integrated software ecosystems 
must undergo that same level of conceptual reinvention, driven by rigorous inquiry and a willingness to redefine established frameworks.

We face the urgent need for a bold, enduring transformation—one that extends beyond technology alone. 
We envision a new kind of agile and robust scientific computing ecosystem built through {\em socio-technical co-design}—i.e, the intentional, integrated development of social and technical components as interdependent parts of a unified strategy.
This approach integrates cutting-edge advances in AI and software—such as large language models (LLMs), scientific machine learning (SciML), reasoning models, and coding agents—with strategies to foster cross-disciplinary team building, training, and collaboration.
Two key themes emerged: 
\vspace{-0.06in}
\begin{titemize}
\item AI is not just another tool—it can dramatically boost scientific productivity, catalyzing new discovery.
Moreover, while demonstrably transformative, AI's evolution is not known.
\item Supporting this transformation in computational science will require a broad, cross-disciplinary community and sustained investment in preparing the next generation and empowering today’s experts.
\end{titemize}

\noindent
With these guiding themes, the workshop focused on three major challenges in 
future 
scientific computing:

\vspace{-0.09in}
\begin{titemize}
\item {Advancing scientific software ecosystems for HPC and AI} of today and tomorrow, 
while ensuring results remain scientifically valid. 
\item {Fostering collaboration} among scientists, AI experts, research software engineers, educators, community leaders, and even AI systems themselves—bridging traditionally siloed roles and domains. This includes private-sector partners, 
key to translating scientific research into real-world applications. 
\item {Rethinking training and workforce development}, to help today’s and tomorrow’s researchers adapt to rapid changes in tools and technology.
\end{titemize}

The workshop identified 
research directions and community actions to address these challenges. These include 
building modular, trustworthy, AI-powered scientific software systems; 
creating frameworks where humans and AI systems can develop and validate code together;
overhauling training pipelines to reflect the pace and nature of modern computational science;
promoting responsible innovation with guidelines for AI use; and
launching pilot programs and partnerships that test new ways of learning and working.

In short, the workshop laid out a vision for the future: one where AI, software, hardware, and human expertise work hand in hand to accelerate scientific progress, broaden access, build the workforce, and preserve the integrity of the scientific method. This report marks the beginning of a long-term, community-driven effort to turn that vision into reality.

The workshop gave us a fresh, encouraging perspective. 
We must recognize that—more than ever—the 
future of science depends on more than computational power; it depends on how we embed human insight within intelligent systems. When we align AI’s  
capabilities with the creativity, intuition, and  
discernment of researchers, we open new pathways for discovery. Purposeful integration of technical and societal dimensions creates a living system—one that learns, adapts, and accelerates progress where it matters most.

\clearpage
\pagenumbering{arabic}
\section{Significant Challenges Facing Computational Science}
\label{sec:intro}

Throughout history, the most profound  technological advances have challenged society’s capacity to adapt and respond at scale. 
Artificial intelligence (AI), while still immature, is 
already poised to influence nearly every facet of human endeavor~(see, e.g., \cite{anthropic2025Sept}). 
Similarly, while its full impact remains uncertain, AI's disruptive and beneficial potential in science and engineering is clear.
In turn, how we integrate AI across science and engineering 
may shape future global leadership.

Scientific discovery through computing underpins
U.S. innovation, economic growth, and national security.   
Over the past fifty years, advances in computing have revolutionized how scientists learn, experiment, and theorize. 
High-performance computing (HPC)---including modeling and simulation, data analytics, machine learning, and AI---is a necessary means of discovery and innovation in essentially all areas of science, engineering, technology, and 
society~\cite{hendrickson2020ascr,GroppHarrisonEtAl2016,KeyesTaylor2011,siam-cse18,NSF-OAC-Blueprint2021,siam-futurecse2025}. 
Today, that pursuit faces formidable
challenges. 
As scientific data, modeling, simulation, and AI grow in volume, complexity, and strategic value, they have become central to the competitiveness of the nation’s 
industrial and manufacturing sectors such as automotive, aerospace, materials, and others. 
To remain a global leader, the U.S. must act with urgency and intention to elevate its scientific research infrastructure—especially through  scientific software ecosystems that integrate AI and advanced computing capabilities.
This is no longer a question of optional
investment; it is a foundational requirement for ensuring our
socioeconomic prosperity, maintaining global competitiveness,
strengthening national security, and securing a 
higher quality of
life for future generations. 

Meeting this challenge requires broad national commitment and coordination. 
The scientific computing community cannot transform alone. 
Government, academia, national labs, and the private sector must invest together in the technical, institutional, and human systems that advance science. 
Scientific computing must evolve not only in tools and platforms but also in collaboration, governance, and talent development.

To meet this challenge, we must develop and adopt  innovative approaches to
the integration and application of scientific computing, learning from the past and preparing for the
future. Scientific progress now hinges on our ability to integrate
AI not only into our research methodologies
but also into the underlying software ecosystems that power them~\cite{DOE-WorkshopReportAI4Science2023,sssdu-workshop-report2023,NAIRR-pilot,nairr-software-workshop-report2025,nas-nnsa2023,AI-for-science2020}. This
integration must be safe, reliable, and agile, demanding
a new generation of scientific software tools that are robust,
scalable, and adaptable to rapidly evolving computing
platforms—not retrofits, but purpose-built ecosystems capable of accelerating discovery in areas such as materials design, energy, biology, astrophysics, manufacturing, and security.

Equally important is the human dimension. Scientific advancement
depends not only on tools and technology but also on the people who create and use
them. We must cultivate a workforce of 
researchers and technologists who are proficient in
developing and using
emerging software ecosystems and AI-enabled research
methodologies~\cite{ascac-workforce2014, ASCAC_ReportGiles2020,web-supercharging-americas-ai-workforce,siam-futurecse2025,ASCACCSGF2025}. These 
people must be empowered to collaborate across disciplines, think
computationally, and adapt rapidly
to evolving technologies. Building this talent pipeline
is not merely about education—it is about building the intellectual
foundation for the nation’s scientific enterprise. 

This is a multidimensional challenge—requiring urgent, coordinated action. We must advance technology, reimagine software, and cultivate a highly skilled workforce simultaneously. 
The United States aspires to remain a global leader in science and innovation~\cite{VAST-final-report2025}, and 
progress in scientific computing is key to that vision. 
Over the coming decades, state-of-the-art science will increasingly rely on the effective, pervasive, and responsible use of AI, placing new demands on our tools, methods, and practices. 
A key enabler of this transformation will be the development of next-generation scientific software that is not only AI-integrated but also scalable, sustainable, and trustworthy. 
But the transformations needed for effective scientific progress are too large and pervasive to be done piecemeal by independent actors.
The scope and urgency of the challenges demand coordinated, community-driven action across institutions, disciplines, and sectors—and ultimately, across national boundaries, while keeping the flexibility required for the dynamic evolution of scientific computing ecosystems. While these early efforts may focus on national coordination, lasting progress will require globally-connected community efforts.

\subsection{Workshop objectives}
\label{sec:workshop-objectives}

To address these critical needs, the workshop {\em Next-Generation Ecosystems for Scientific Computing: Harnessing Community, Software, and AI for Cross-Disciplinary Team Science} was held in Chicago, IL, during April 29 - May 1, 2025.\footnote{Appendices \ref{sec:workshop-description}, \ref{sec:workshop-participants}, and \ref{sec:workshop-agenda} provide the workshop description, list of participants, and agenda, respectively.} 
The workshop convened a broad group of experts spanning HPC, AI, computational science, domain sciences, applied mathematics, software engineering, cognitive and social sciences, and community development. 
Participants collectively examined how we can co-design next-generation scientific software ecosystems to enable cutting-edge research, foster cross-disciplinary collaboration, and scale effectively across institutions and infrastructure.
The workshop introduced the methodology of {\em socio-technical co-design for next-generation scientific computing}, explained in Section~\ref{sec:socio-technical-co-design}, as a complement to insights from other forward-looking community reports (e.g., \cite{nairr-software-workshop-report2025,DOE-WorkshopReportAI4Science2023,sssdu-workshop-report2023,ASCR-facilities2024,software-engineering-future-sei2021,DOE-envisioning-science2050}). 
By intentionally blending technical and community factors, the workshop aimed to shape a future where thriving, cross-disciplinary communities drive the next wave of scientific discovery, with high-quality scientific software as a keystone of sustained collaboration and scientific progress. 

\begin{figure}[htb]
\begin{center}
\vspace{-0.20in}
\includegraphics[width=0.56\textwidth] {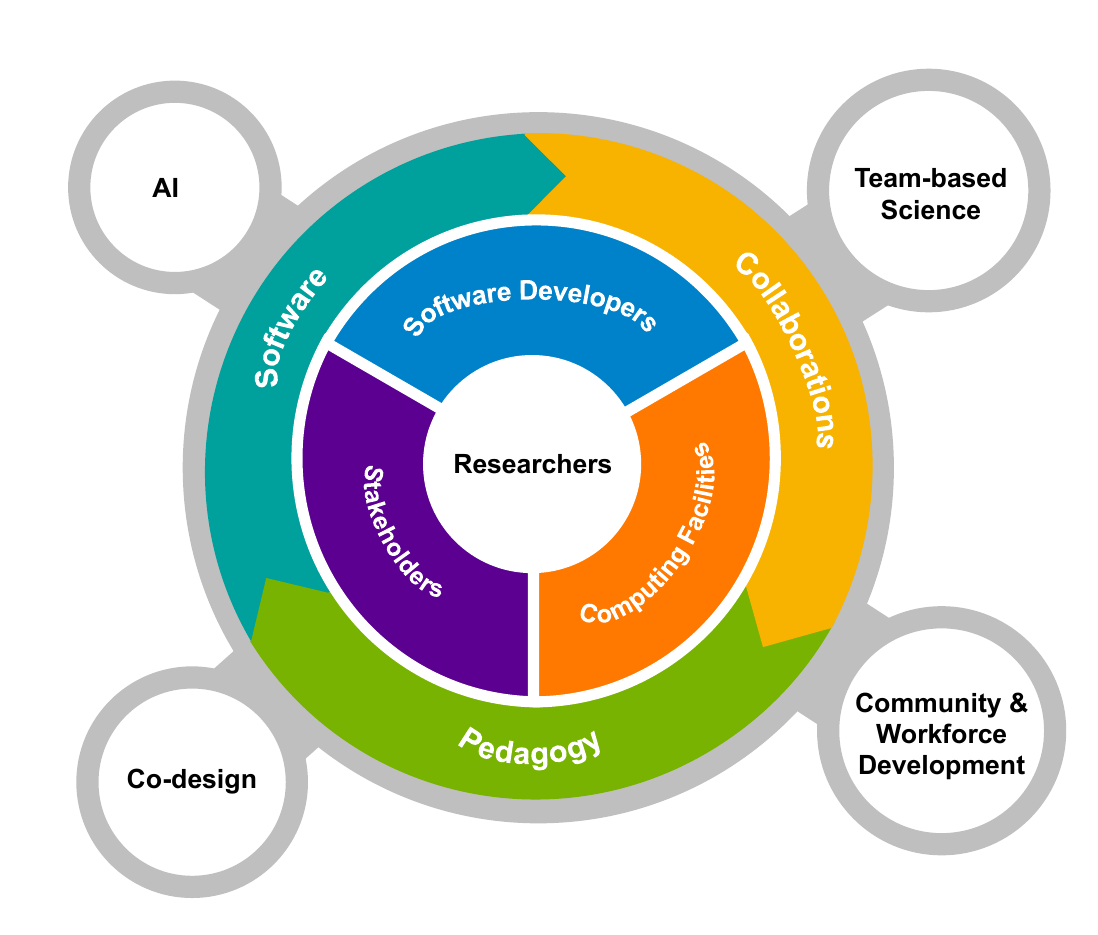}
\vspace{-0.20in}
\caption{Toward next-generation ecosystems for scientific computing: As motivated by the needs of team-based science in an AI-driven future, we must advance software, cross-disciplinary collaboration, and pedagogy (outer ring of this figure) through co-design, all while developing workforce and community. This work demands strong collaboration among researchers (in science domains, applied math, computer science, HPC, AI, etc.), software developers, stakeholders, and computing facilities (inner rings of this figure).} 
\label{fig:ecosystems-for-scientific-computing}
\end{center}
\end{figure}

\label{def:ecosystems}
\paragraph{Scientific computing ecosystems.} In the context of this report, {\em ecosystems for scientific computing} are dynamic socio-technical systems made up of people, technologies, infrastructure, institutions, workflows, and cultural practices that collectively support the development and evolution of scientific computing. 
Ecosystems are role- and context-dependent—shaped by the interactions among stakeholders, researchers, developers, users, institutions, and AI systems. A successful ecosystem integrates technical components (such as software libraries, AI technologies, and computational platforms) with social dimensions (such as training, governance, and collaboration norms) to enable scientific progress. 
We believe that such ecosystems will emerge
through the deliberate co-design of software platforms, scientific tools, community structures, and educational models—systems that support rapid change and are coordinated through well-designed, cooperative approaches rather than centralized control. Figure~\ref{fig:ecosystems-for-scientific-computing} illustrates aspects of these relationships. 

\subsection{Potential impact}

\begin{figure}[bht]
\vspace{-0.8in}
\begin{center}
\includegraphics[width=0.99\textwidth] {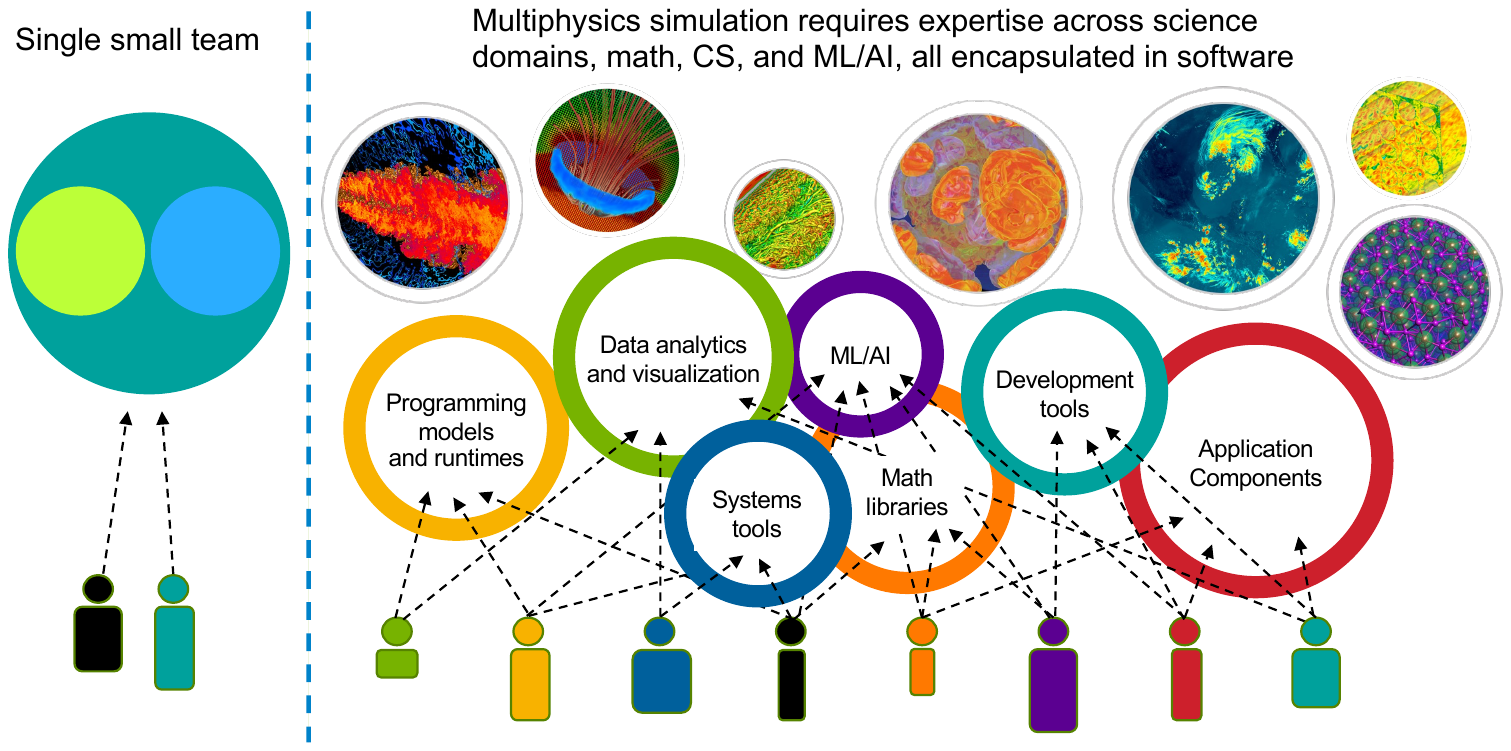}
\vspace{-1.1in}
\caption{Simulations in advanced scientific computing (including materials science, astrophysics, nuclear energy, biology, engine design, weather prediction, and batteries, as represented by the science images of this figure) require collaboration across science domains, applied mathematics, computer science, ML/AI, and more, where high-quality software is a primary means of encapsulating expertise for use by others. 
Due to increasing science challenges and complexity, no longer can a single small team (shown on the left-hand side) independently develop all required functionality.  Instead, reusable libraries and tools (shown on the right-hand side), developed by teams whose expertise spans across various topics, provide key functionalities that serve many applications. Dashed lines indicate multiple areas of work per person (e.g., the red-colored person in this diagram contributes to application components, development tools, and ML/AI capabilities).}
\label{fig:ecosystems-collaboration}
\end{center}
\end{figure}

Dynamic computational workflows, encompassing 
modeling, simulation, data analytics, and emerging AI approaches, as enabled by the instrumental building blocks of scientific software, are
key to next-generation science~\cite{DOE-WorkshopReportAI4Science2023,pyzer-knapp2022,Boiko2023,NASEM26894_2024,Bilodeau2024,spangenberg2025msiFlow}. 
Ultimately, the promise---and eventual impact---of 
addressing the urgent challenges introduced above rests on the
ability of distributed, cross-disciplinary teams to effectively
integrate varied knowledge, perspectives, policies, requirements, and
methods toward producing innovative solutions for scientific computing. 
Figure~\ref{fig:ecosystems-collaboration} illustrates the cross-disciplinary collaboration that is typical in advanced computational science---broadly considering simulations in physics, chemistry, materials science, biology, and so on,\footnote{For example, see the wide range of scientific applications within the recent DOE Exascale Computing Project (ECP)~\cite{ecp-website,alexander_exascale_2020}.} 
which leverage heterogeneous computing architectures.   
These scientific codes build on 
low-level programming models and runtimes, libraries for math and visualization, tools to understand and optimize performance, 
emerging ML/AI technologies (such as distributed training frameworks, MLOps pipelines, and federated learning systems), application-specific components, and more.\footnote{For example, E4S~\cite{E4S-web} provides a foundational HPC-AI software ecosystem for science~\cite{HerouxEtAl2024}, 
including ECP libraries and tools~\cite{SWEcosystems:NCS2021,ECP-software-technologies2024,TransformingScienceThroughSoftware2024} as well as popular AI packages.} 

\vspace{-0.1in}
Opportunities for advances via AI technologies abound, including code generation, automated design of experiments, AI agents, and real-time analysis during simulations. 
For example, a materials science team using AI to predict crystal structures needs seamless integration between calculations of density functional theory, ML training, and HPC scheduling, while building on programming models, systems tools, and so on.
Likewise, an astrophysics team using HPC and AI to study supernova explosions needs a multiphysics simulation code with shock hydrodynamics and nuclear burning, which might rely on math libraries for adaptive mesh refinement and on AI for parameterized physics models such as flame and equation of state. 

Complexity arises due to heterogeneity across and within 
the various components constituting next-generation scientific computing ecosystems. 
This heterogeneity exists at many levels, for example in
(1) hardware (e.g., CPUs, AI accelerators, quantum); (2) software
(e.g., multitude of programming languages, environments, functionalities); (3)
algorithms (e.g., mathematical, computer science, statistical, AI,
domain science specific); (4) precision of operations (e.g., mixed vs.\ double precision, quantization in AI models); (5) data (e.g., storage, movement,
processing, analysis, visualization); (6)~computing environments
(e.g., leadership computing facilities, federated, cloud, edge, quantum); and (7) the workforce (e.g., 
researchers, developers and users of software, domain scientists, 
management, stakeholders) or, more generally, cross-disciplinary teams, which in
next-generation scientific computing may eventually include
AI agents alongside humans
\cite{lu2024aiscientistfullyautomated, 
Swanson2024VirtualLabAgents,ghareeb2025robinmultiagentautomatingscientific}.
This report considers the integration
of AI into scientific software development as a transformative
opportunity to accelerate discovery while addressing long-standing
challenges in code quality, sustainability, and accessibility; 
other documents consider complementary topics (see, e.g., \cite{heterogeneity2018,data2021,quantum2023}).
AI's promise lies not just in automating routine coding tasks but also in 
fundamentally reimagining how scientific software ecosystems evolve. 

Collaboration is
a primary mode of work in scientific computing, yet its potential is only beginning to be fully realized because of
the complexity of cross-disciplinary
research.  
The advent of AI, with its potential to influence 
human activities related to software development and scientific
discovery, adds a transformative new dimension to this challenge.
Next-generation ecosystems in scientific computing 
offer the promise of 
integrating broad domain expertise, methodological
approaches, and AI-driven tools to bridge gaps between theoretical
principles and practical implementations. These advances can enable more
robust, maintainable, and scientifically accurate software, thereby accelerating scientific discovery. 

We can build on experiences of innovative programs such as the U.S.\ Department of Energy's (DOE's) 
Exascale Computing Project~(ECP; \cite{ecp-website,ecp-kothe-lee-qualters-2019}), 
Scientific Discovery through Advanced Computing (SciDAC) program~\cite{scidac-website}, and 
Computational Science Graduate Fellowship (CSGF) program~\cite{CSGF2021, csgf-website};
the National Nuclear Security Administration's (NNSA's) 
Predictive Science Academic Alliance Program~\cite{nnsa-psaap-website}; the
National Science Foundation's (NSF's) 
Science and Technology Centers~\cite{nsf-stc-website} and 
Engineering Research Centers~\cite{nsf-erc-website};
Sustainable Research Pathways~\cite{SRP-website}, and others.
These initiatives have
pioneered bold approaches for cross-disciplinary collaboration at scale 
and for cultivating both breadth across disciplines and depth within individual fields—countering 
traditional academic silos, as required for advances in 
scientific computing. 

\subsection{Socio-technical co-design for next-generation scientific computing}
\label{sec:socio-technical-co-design}

\begin{figure}[hbt]
\begin{center}
\vspace{-0.1in}
\includegraphics[width=0.99\textwidth,trim={0.0cm 3.3cm 0.0cm 3.3cm},clip]{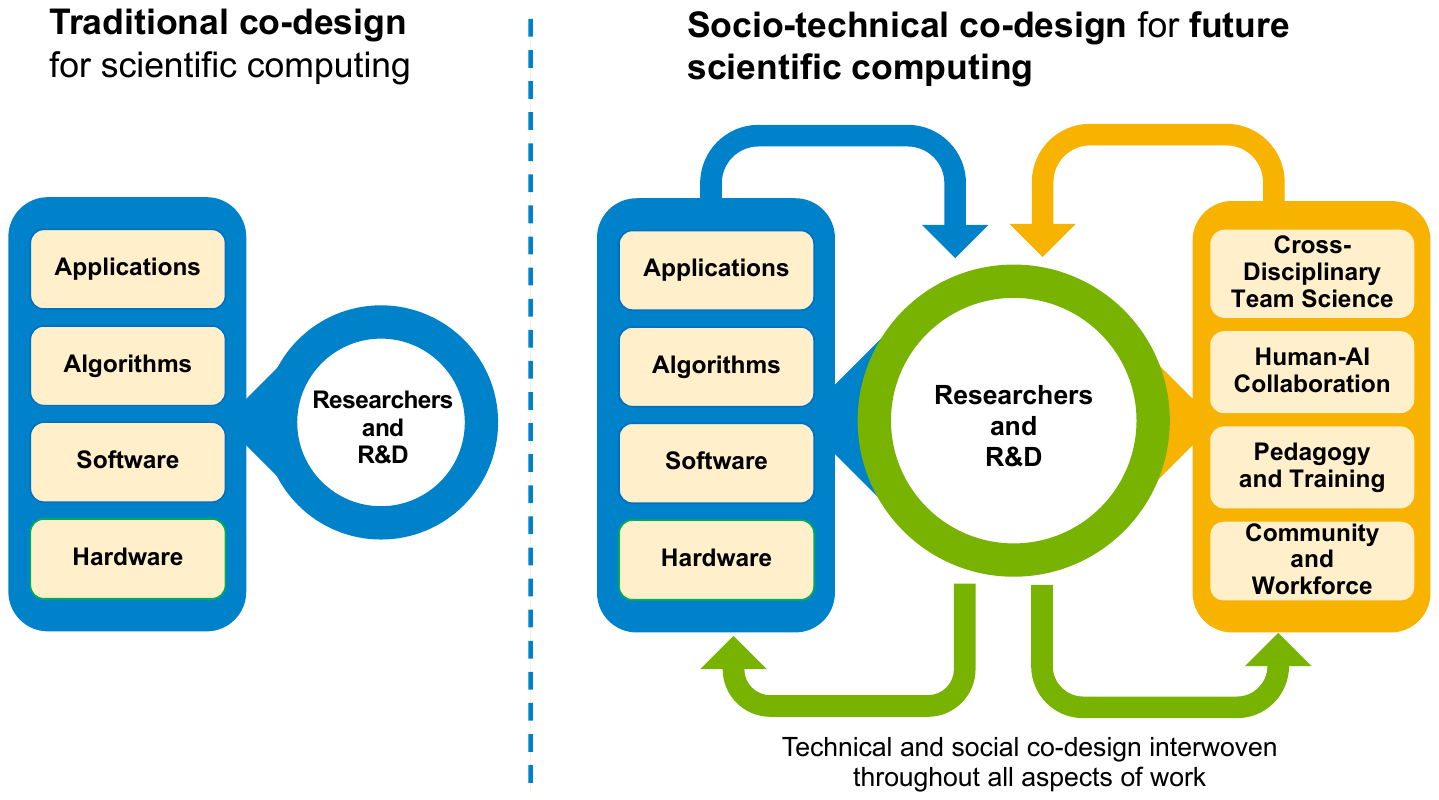}
\vspace{-0.2in}
\caption{Socio-technical co-design for next-generation scientific computing intentionally interweaves technical and social elements throughout all aspects of work, while closely coupling cycles of R\&D innovation between computing technologies and driving applications.  
By ensuring that technical and social dimensions are addressed as tightly coupled elements of a unified, forward-looking strategy, this holistic approach 
accelerates transformative impact across wide-ranging application domains.
}
\label{fig:socio-technical-co-design}
\end{center}
\end{figure}

To tackle these urgent, complex, and intertwined challenges, the workshop introduced a novel methodology of {\bf socio-technical co-design} for scientific computing, illustrated in Figure \ref{fig:socio-technical-co-design}.  
The concept of co-design is widely used in many fields, referring generally to 
collaborative approaches to designing solutions by involving multiple stakeholders (especially those who will be using or are impacted by the outcomes) throughout the design process.
For example, in the human-computer interaction community, co-design is strongly associated with participatory design.
Co-design in HPC and AI has been critical to the design and implementation of contemporary computer architectures, applications, algorithms, and software, considering their interrelationships (as shown in the left-hand side of this figure)~\cite{co-design-workshop-report2022,IJHPCA-codesign}.

Building on these co-design experiences, the workshop introduced the broader approach of 
socio-technical co-design, which embraces the intentional and integrated development of both technical components (e.g., software, AI, infrastructure) and social components (e.g., teams, institutions, practices, training) of scientific computing ecosystems. This expanded approach ensures that social and technical dimensions are addressed not in isolation but as tightly coupled elements of a unified, forward-looking strategy.

\paragraph{Report structure.} We organize insights from the workshop across three complementary axes
that undergird scientific computing and urgently require
better understanding and advancement to prepare for AI’s growing role in shaping scientific discovery (see Figure~\ref{fig:report-structure}): 

\begin{figure}[b]
\begin{center}
\vspace{-0.6in}
\includegraphics[width=0.95\textwidth,trim={0.0cm 6.5cm 0.0cm 5.5cm},clip]{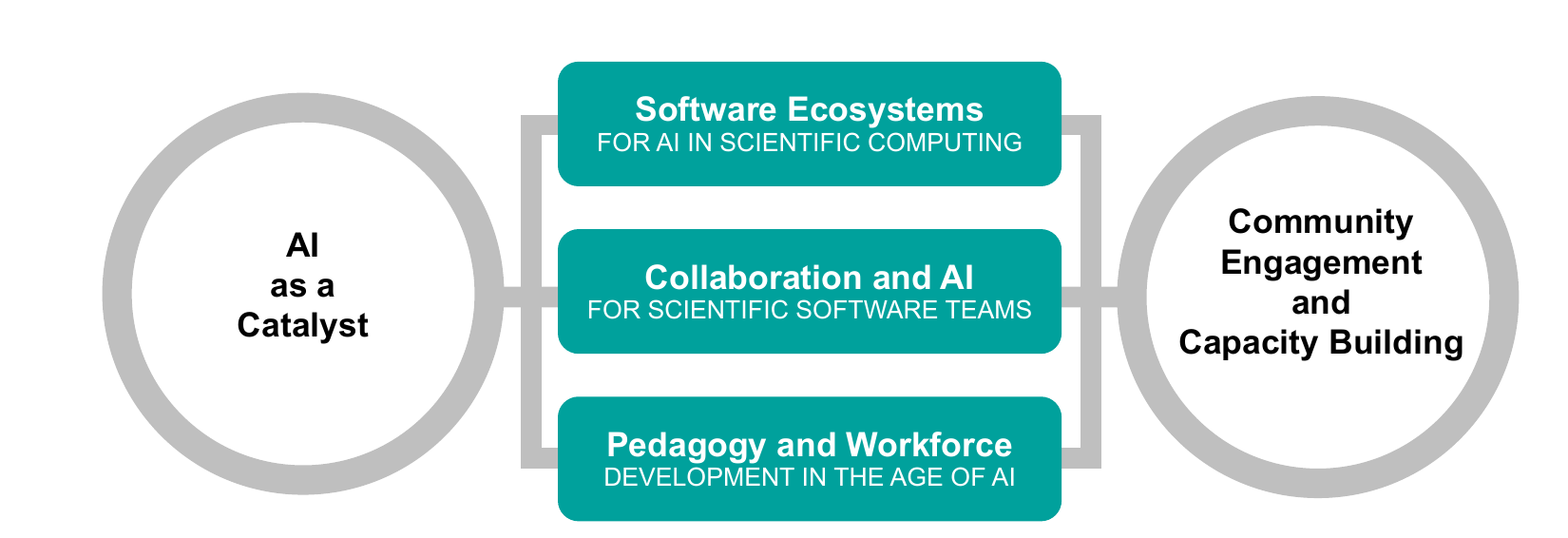}
\caption{Report structure.}
\label{fig:report-structure}
\end{center}
\end{figure}

\begin{titemize} 
\item Software ecosystems for AI in scientific computing
\item Cross-disciplinary collaboration and AI for scientific software teams
\item Pedagogy and workforce development in the age of AI
\end{titemize}

\newpage
\noindent
Two crosscutting topics permeate all of these:  
\begin{titemize}
\item AI as a catalyst to advance software productivity, collaboration, and pedagogy
\item Community engagement and capacity building for next-generation scientific computing
\end{titemize}

\noindent
These domains are deeply interconnected, requiring co-design of technical and social systems. 
Addressing them holistically presents an opportunity to foster robust and scalable research ecosystems capable of driving scientific discovery in the coming decade.
We next delve into these challenges, including crosscuts and interdependencies, and we introduce priority research directions as well as actionable community strategies.  

\section{Breaking Down the Challenges}
\label{sec:challenges}

As scientific computing moves toward AI-driven, cross-disciplinary, community-integrated ecosystems, significant challenges arise in aligning technical tools with human needs and institutional capabilities. 
Sections~\ref{sec:challenges-software-ecosystems} through~\ref{sec:challenges-pedagogy-community} elaborate on challenges in the complementary topics introduced in Section~\ref{sec:intro}, while Section~\ref{sec:challenges-crosscutting} discusses crosscutting issues and interdependencies.

\subsection{Software ecosystems for AI in scientific computing}
\label{sec:challenges-software-ecosystems}
  
We define a software ecosystem for
AI-enhanced scientific computing as the entire software infrastructure
that supports all computational steps involved in scientific
discovery, ranging from  system software at the lowest level to
inference engines at the highest level.  The inference engines rely
upon data obtained from simulations, observations, archives, and other
sources of scientific data.  Simulation and analysis tools in turn
depend upon middle layers of the software stack such as math and
visualization libraries and workflow management tools.  Tools to
analyze, predict, and optimize performance, energy, and cost of
simulations are also essential. The software ecosystems must not
only integrate traditional numerical methods and scientific
simulations but also accommodate rapidly evolving AI models, data
pipelines, and heterogeneous computing environments (e.g., CPUs, GPUs,
and quantum systems).  While heterogeneity is not new to scientific computing, at present it
is growing in complexity and ever-changing, raising questions about
CPU-GPU memory transfers, mixed precision training, and quantum-classical hybrid algorithms.
Key questions thus
revolve around the urgency to identify what will be needed to
design, develop, and maintain software that will successfully support
next-generation science conducted under heterogeneous, dynamic, and 
fast-evolving environments, 
while simultaneously building the human enterprise and collaboration protocols. Specific software needs are discussed in the recent Report of the NSF/DOE Workshop on NAIRR Software~\cite{nairr-software-workshop-report2025}, while software engineering challenges are presented in~\cite{software-engineering-future-sei2021}.
Creating extensible and
sustainable ecosystems will require a strategic blend of technical innovation and
community-driven co-design. 

\subsubsection*{Key questions:} 
\begin{itemize}
\item How can we build extensible, traceable, and modular scientific
  software ecosystems that support both legacy and AI-generated
  components, and what degree of intrusion into existing code is
  needed and acceptable? Can AI assist in decomposing monolithic
  scientific codes into reusable components that support ecosystem
  integration? How can AI tools help with refactoring, modernization,
  and documentation of legacy scientific code?

\item How can AI maintain scientific correctness while accelerating
development? Scientific applications must adhere to physical laws,
mathematical principles, and domain-specific constraints. How can we
train AI systems to generate code that is scientifically valid and
numerically stable?

\item What validation frameworks are needed for AI-generated
  scientific code? Traditional software testing approaches may be
  insufficient for scientific applications where correctness extends
  beyond functional requirements to include adherence to scientific
  principles, uncertainty quantification, and reproducibility across
  different computational environments.

 \item  How can standards for AI be designed and integrated into
  software, applications, tools, and reporting  to ensure the
  technologies can be trusted and adopted by different segments of the
  scientific computing community with relative ease?  
\end{itemize}

\subsubsection*{Scientific challenges and opportunities:}

\paragraph{Fragmented ecosystem and interoperability challenges.}
  Scientific software ecosystems consist of wide-ranging
  codebases that already struggle with continually changing hardware
  paradigms and now must also integrate with modern AI frameworks. 
We face challenges in
  bridging traditional scientific computing environments with
  AI-driven development workflows, including deploying trained models in HPC environments and
  integrating scientific data formats with AI frameworks.
Given that data is a critical component of AI-enabled scientific computing, and that computational results are often distributed across teams and siloed in storage systems, seamless access to data remains a major barrier. To support effective collaboration, we need standardized mechanisms for data sharing that address access control, provenance, and metadata.
The lack of standardized interfaces, data
  formats, and integration protocols—especially for
  deeply integrated components like math libraries—further prevents seamless adoption of AI technologies and forces scientists to navigate
  multiple, incompatible systems.
This fragmentation creates
  inefficiencies, requires manual intervention between simulation and
  analysis stages, and limits the ability to leverage AI across entire
  scientific workflows. 
Additional challenges relate to the need for
performance optimization, such as model quantization, pruning, and knowledge distillation, 
as well as
adaptability in dynamic AI landscapes.
 
\paragraph{Reliability and validation gap in AI-generated scientific code.}
A critical challenge 
is the disconnect between AI's current capabilities and the 
requirements of scientific computing for reliability and replicability~\cite{nas-reliability2015,nas-reproduciblity2019}.
Unlike general-purpose software,
scientific applications must preserve physical principles, maintain
numerical stability, and provide reproducible results across different
computational environments. Current AI tools often generate
syntactically correct code that may violate domain-specific
constraints or introduce subtle numerical errors that compound over
long simulations. This point is crucial because natural
language specification is by its nature imprecise. Traditionally, programmers eliminate artifacts of imprecise
specification through iterative verification and
debugging. What can replace or accelerate this iterative convergence to
trustworthy software in its development cycle remains an open
question. This gap is exacerbated by the lack of standardized
validation frameworks specifically designed for AI-generated
scientific code, making it difficult to systematically assess whether 
generated solutions maintain both computational correctness \cite{gokhale2023correctness} and
scientific validity. 
Federated learning introduces additional challenges, including needs for privacy-preserving collaboration across institutions and addressing security threats such as adversarial attacks on scientific AI models and data poisoning.
Additionally, human-in-the-loop and continuous
learning mechanisms will remain essential for aligning AI-generated software
with evolving scientific understanding.

\paragraph{Development methodologies for hybrid modeling/simulation and AI.} While it will be impractical to develop entirely new scientific software ecosystems from
scratch, reuse of existing software in these new modalities is
likely to be challenging. The changes required to code will be 
deeply invasive; therefore, intrusion-aware design will become
critical to minimize disruption when integrating new tools or AI
systems into existing software. There is an ongoing debate in the
scientific community about when and where  traditional numerics should be used,
where emerging scientific machine learning
(SciML~\cite{Baker2019}) methods might replace them, and how the two can most effectively complement one other.
We urgently need to develop (and adopt) suitable
metrics when evaluating SciML  methods against traditional numerical
methods 
\cite{mcgreivy2024weak,brandstetter2025envisioning}.    
Middleware to
assist in the creation and execution of automated computational
workflows will continue to be essential but will now also need
consideration for agentic AI systems \cite{pauloski2025FederatedAgents}. To support scalable infrastructure
growth, such advances require
ecosystem-level funding models, broad coordination, and consortia involving national 
labs, academia, and the private sector.

\subsection{Cross-disciplinary collaboration and AI for scientific software teams}
\label{sec:challenges-teams} 

\begin{wrapfigure}{R}{0.36\textwidth}
\includegraphics[width=0.36\textwidth]{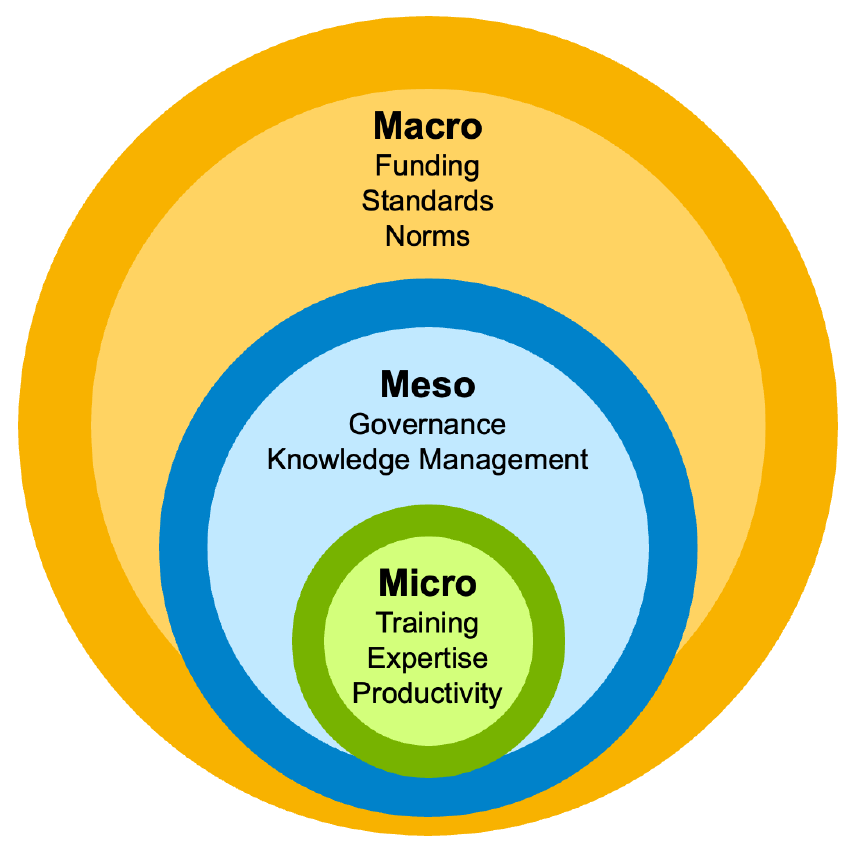}
\vspace{-0.2in}
\caption{Team challenges span throughout macro-, meso-, and micro-level factors.}
\label{fig:teams-macro-meso-micro}
\end{wrapfigure}

Because the physical world is not divided into specific domains, those who seek to understand it must also eliminate unnatural boundary lines dividing their discourse.  
Computational science in general, and high-performance computing in particular, have pioneered advances in cross-disciplinary teams 
combining computing with science, engineering, and mathematical foundations.
Such scientific teamwork, in which an interdependent set of group members share a common purpose, performance goals, and mutual accountability, has grown significantly to meet the needs of organizations and tackle formidable problems \cite{Wuchty2007, Williams2008,nas-teams2015,nas-teams2025}.
However, the growing complexity of challenges in scientific computing demands 
increasing collaboration among researchers and potentially AI agents---where software is a primary means of collaboration. 

Cross-disciplinary teams in scientific computing face communication and knowledge integration challenges due to diverse vocabularies, standards, and methodologies~\cite{Fiore2008, nas-teams2025}.
These challenges are associated with the structure and norms of scientific research; the proliferation of data and computational technologies; the complexity of interactions; and the knowledge, skills, and attitudes present in teams.
The rapid evolution of AI technologies complicates each of these issues and also introduces a unique set of challenges.
A multipronged strategy addressing macro-, meso-, and micro-level factors (see Figure~\ref{fig:teams-macro-meso-micro}) that influence collaboration in scientific computing is of paramount importance; consequently, we structure the remainder of this section according to these factors. 

\paragraph{Macro-level factors (science and society): Structures and norms for collaboration} 

\paragraph{Key questions:}
\begin{itemize}
 \item How can existing funding structures be adapted to encourage greater and optimal interconnectedness in the scientific computing community? What new mechanisms, and possibly new sources for funding, are needed to encourage and maintain collaborations in scientific computing?
 \item How can existing norms for recognition and credit be adapted to encourage greater and optimal sharing and cross-disciplinary collaboration in the scientific computing community? What novel norms for recognition and credit in cross-disciplinary research are needed to overcome the limitations of siloed, unidisciplinary structures in academia?
 \item How can standards for AI technologies be designed to facilitate rather than constrain collaboration and innovation? 
\end{itemize}

\paragraph{Scientific challenges and opportunities:}
 
The structures and norms that shape or otherwise enable scientific research are often at odds with the approaches and goals that characterize cross-disciplinary collaboration. 
These include the availability and distribution of funding (often insufficient for the full scope of multi-institutional, cross-disciplinary multiteam systems \cite{nas-teams2025} required for large-scale computing projects). 
Another macro-level challenge is norms for recognition and credit, which sometimes neglect or minimize the work of various types of contributors in projects, resulting in decreased motivation, satisfaction, and productivity.
Reward systems need to adapt accordingly, reflecting a shared value system for all components of the evolving landscape, including the development and maintenance of the underlying supports such as software “plumbing,” project management, and communities that support collective needs.

\paragraph{Meso-level factors (organizations, interactions): Models of governance and knowledge management} 

\paragraph{Key questions:}
\begin{itemize}
  \item What forms of governance are most beneficial for achieving fair and defensible decision making in cross-disciplinary collaborations in scientific computing?
  \item How can these governance models enable effective collaboration, accelerate innovation, and help ensure that all relevant perspectives are represented? What resources and tools are needed to enable the adoption of these governance models in multiteam collaborations? 
  \item What approaches to knowledge management are beneficial for fostering trust, mitigating knowledge loss, and reducing barriers to contribution and collaboration? 
\end{itemize}

\paragraph{Scientific challenges and opportunities:}

The complexity of collaboration in scientific computing is associated with the number of stakeholders and the differences in their requirements, approaches, and goals. 
Effectively managing this complexity is vital for the quality of collaboration processes and outcomes, including communication, coordination, decision-making, problem solving, performance, and knowledge transfer~\cite{nas-teams2025}.
Access to knowledge must become ubiquitous, fair, and democratic, taking advantage of AI and cloud-based approaches.
This challenge introduces an opportunity to envision models of governance and knowledge management that support distributed collaboration across institutions, ensuring shared responsibility for next-generation ecosystems in scientific computing. 

\paragraph{Micro-level factors (teams and constituent members): Training for expertise and productivity} 

\paragraph{Key questions:}
\begin{itemize}
  \item How can collaboration among various actors in scientific computing (including national laboratories, universities, and the private sector) be leveraged to design and provide opportunities for training in cross-disciplinary research models and teamwork competencies?
  \item How can human-centered strategies and AI technologies be designed to support the acquisition of {\em taskwork} competencies (e.g., knowledge, skills, and abilities required for writing high-quality software, evaluating and selecting software, modifying and adapting software) 
  as well as {\em teamwork} competencies (e.g., knowledge, skills, and abilities required for effective communication, coordination, learning) in scientific computing?
  \item How can human-centered strategies and AI technologies be designed to minimize workload, reduce miscommunication, increase efficiency, and augment the capabilities of individuals and teams in scientific computing projects? How can human-centered strategies and AI technologies be designed to enhance team formation and team development across multiteam collaborations? 
   \end{itemize}

\paragraph{Scientific challenges and opportunities:}

Cross-disciplinary research teams face significant challenges in communication, which in turn can reduce their capacity for effectively utilizing collaborative member expertise, developing shared knowledge structures (i.e., shared mental models and transactive memory systems), and integrating their varied knowledge \cite{Fiore2008}.
Because cross-disciplinary research demands the integration of the varied, specialized expertise present in a team, its members must be adept in communication and exhibit openness toward one other---competencies and dispositions that influence the team’s expertise utilization and knowledge integration. 
Furthermore, individuals in these teams must develop expertise in topics deemed essential for scientific computing (e.g., reproducibility~\cite{nas-reproduciblity2019}).
Both training and research are thus needed on attitudes, behaviors, and cognition that can enhance cross-disciplinary collaboration in the scientific computing community.
Also needed are advances in
AI-specific collaboration tools, such as shared Jupyter environments, collaborative model development platforms (GitHub for models), and automated code review systems for AI-generated code. 
For example, domain scientists need to be able to validate AI-generated finite difference schemes without understanding all details of the underlying transformer architecture.

\paragraph{Multilevel factors: Trust in teammates, software, and AI.} 

Achieving an appropriate level of trust in teammates, software, and AI represents an additional significant challenge.  
Developing and maintaining trust across the macro, meso, and micro levels of collaborative scientific computing requires insightful leadership, time, repeated interaction, and intervention when trust is lost. 
A challenge is devising alternative means for achieving trust within an ecosystem---minimizing the workload required to develop trust and manage uncertainty associated with information 
(e.g., by leveraging wisdom of the crowds and collective intelligence).
Also needed are trust protocols for AI agents, for example for AI agents to submit computational jobs to leadership-class facilities.

\subsection{Pedagogy and workforce development in the age of AI}
\label{sec:challenges-pedagogy-community} 

Pedagogy and workforce development in the computing sciences
have always been challenging because of the need to understand and
communicate ideas that span multiple disciplines. In academia an added
challenge has been the difficulty of fitting in 
cross-disciplinary training in traditional department
structures~\cite{siam-cse18,siam-futurecse2025}. Mission-driven research labs have been the natural
homes for such efforts. With the advent of AI and growth of
computational science---spanning beyond traditional roots in the
physical sciences to include life sciences, social sciences, and
virtually all aspects of science and society---there is more opportunity and urgency to integrate cross-disciplinary training. 
Many science and engineering 
departments will need to 
extend their curricula to include developing expertise in exploiting powerful computing platforms, including AI, in much the same way they have trained students in experimental methods in the past. 

\paragraph{Key questions:} 
\begin{itemize}
\item How do we prepare the scientific workforce for an AI-augmented
  future? As AI tools become more sophisticated, the roles of domain
  scientists and software engineers will evolve. Critical questions
  include the following: What fundamental skills remain essential? What specific AI/ML concepts are essential? How do we
  maintain critical thinking and domain expertise when AI handles
  routine tasks? How do we ensure that the next generation can validate and
  improve upon AI-generated solutions?
How can we ensure that AI models and datasets reflect a wide range of perspectives grounded in real-world contexts?

\item 
  How can strategies be developed to train and grow workforce
  communities when the requirements are unknown and rapidly changing?
  Moreover, how can we broadly enhance and evolve education to foster
  critical thinking, extend access, expand cross-disciplinary
  collaboration and communication, and create ubiquitous and
  democratic access to and implementation of AI for computational science? 
  How can we effectively address community building and workforce development among various stakeholder groups?

\item How can we build training catalogs for community-wide use, including all relevant perspectives, so
  that we can broaden the pool of scientists and engineers engaged in
  computational science?
\end{itemize}

\subsubsection*{Scientific challenges and opportunities:}

\paragraph{Workforce development and knowledge transfer crisis.}
Perhaps the most profound challenge identified in the workshop is the
degree of uncertainty in how to prepare the future workforce for the
disruptive changes underway in computing for science. 
As AI tools become more capable, there is growing concern
about creating a generation of scientists who can use AI-generated
code but lack the fundamental understanding needed to validate, debug,
or extend it. This situation creates a paradox: AI tools designed to democratize
software development may inadvertently create new barriers to deep
scientific understanding. Simultaneously, the rapid pace of AI
advancement makes it difficult for educational institutions and
training programs to update curricula as technologies and practices change, while the shift toward
AI-assisted development raises fundamental questions about what
constitutes essential knowledge in the age of artificial
intelligence. 
Building curricula and the workforce for rapidly changing technological ecosystems requires new instructional and workforce development paradigms, with higher reliance on apprenticeship models, 
including the social (group) learning that is supported by communities of practice~\cite{LaveWenger1992}.

\paragraph{Strategies to bridge across disciplines.}
The development of integrated, cross-disciplinary curricula in scientific computing has been a slow and uneven process; despite decades of progress, many academic institutions still lack programs in computational science and engineering. 
As next-generation scientific challenges increasingly demand complex, team-based approaches that span multiple disciplines, it is essential to accelerate efforts to design curricula that bridge scientific domains, computational methods, and technological modalities. 
To support this shift, there are growing opportunities to leverage cloud-based AI platforms and containerized environments that enable consistent, scalable training across diverse institutional contexts.
We also need to incorporate critical professional competencies often missing from current curricula, such as project management, agile development practices, effective communication, and collaborative teamwork.
 
\paragraph{Gaps in workforce readiness and training.}
One of the most persistent challenges in computational science is 
the limited pool of qualified candidates for recruitment—a problem compounded by the scarcity of formal academic programs in the field. Although various approaches have been attempted, they have not fully addressed the underlying gap in training pathways. 
Apprenticeship models for hands-on learning have shown great promise as a complementary approach to formal instruction. 
These are typically well-supported by associated communities of practice, where practitioners can continue to share knowledge and tactics in an ongoing, informal manner that supports continuous learning~\cite{WengerEtAl2002}.
Another promising approach is to shift from solely recruiting individuals with comprehensive subject-matter expertise to also recruiting based on potential, followed by targeted, domain-specific training. 
However, this strategy introduces a new set of challenges: developing high-quality training content, keeping it up to date with evolving technologies, and delivering it at scale. These tasks require sustained resource investment and strong coordination across the broader scientific and educational communities.

\subsection{Crosscuts and interdependencies}
\label{sec:challenges-crosscutting}

To fully understand the challenges and opportunities identified in Sections~\ref{sec:challenges-software-ecosystems} through \ref{sec:challenges-pedagogy-community}, it is essential to consider two crosscutting themes that emerged consistently throughout the workshop: (1) the transformative role of AI in improving scientific software productivity, collaboration, and pedagogy; and (2) the central importance of community engagement and capacity building in enabling and sustaining these transformations in the computing sciences. 
These crosscutting themes intersect with all three core challenge areas and are foundational to building resilient next-generation scientific computing ecosystems.

\paragraph{AI as a catalyst for scientific productivity and sustainability.}
Artificial intelligence is not just a topic unto itself—it is a force multiplier across software ecosystems, collaboration frameworks, and pedagogical approaches. 
Section~\ref{sec:challenges-software-ecosystems} discusses the need for AI-integrated software ecosystems that are modular, traceable, and scientifically valid. 
AI can play a key role in this work by helping to advance software productivity and sustainability, such as enabling intelligent code generation, automating routine tasks, and supporting decision-making across heterogeneous computing environments. 
AI also presents new challenges in terms of verification, trust, and intrusion-aware integration.

AI can promote collaboration by acting as a bridge across disciplinary boundaries.  As discussed in Section~\ref{sec:challenges-teams}, 
AI can support shared understanding by translating domain-specific terminology, managing distributed workflows, and facilitating knowledge retention. 
AI systems can also act as teammates—assisting with code reviews, optimizing team formation, and recommending learning pathways. 
These roles require careful governance to ensure they support rather than hinder team cohesion and trust.

As explored in Section~\ref{sec:challenges-pedagogy-community}, AI can also transform pedagogy by offering new modalities of instruction, such as AI tutors, generative feedback, and adaptive curriculum planning. 
However, these benefits must be balanced against the risk of over-reliance, which could erode critical thinking and domain expertise. 
AI-enhanced education systems must therefore be grounded in robust, fair frameworks and designed to complement (not replace) human mentorship and community support.

\paragraph{Community engagement and capacity building as a strategic imperative.}
As AI transforms the technical foundations of scientific computing, the long-term success of that transformation will depend equally on the strength of the communities that adopt, adapt, and sustain it.
Section~\ref{sec:challenges-software-ecosystems} highlights that successful integration of AI into software ecosystems 
requires more than technical proficiency—it demands fluency in interdisciplinary thinking, sound judgment, and collaborative practices. Building this capacity across communities is essential for maintaining and evolving shared software infrastructure.

As discussed in Section~\ref{sec:challenges-teams}, the human dimension of collaboration  is central. Scientific progress increasingly depends on teams with complementary skills, diverse perspectives, and broad participation. 
Capacity-building strategies—such as training programs, fellowships, internships, apprenticeships, and institutional partnerships—should be designed to foster cross-disciplinary fluency, resilience, and leadership.
Community norms, as well as opportunities for professional development, must evolve to recognize and reward all contributors, including those in roles that have historically been undervalued.

Section~\ref{sec:challenges-pedagogy-community} underscores the need to rethink education and community building considering these dynamics. 
Curriculum innovation, community-led training models, and broad access to computational and human resources are critical. 
Well-supported communities create networks of practice that encourage and reward innovation and different perspectives, share lessons learned, and mentor the next generation of scientists, developers, and educators.

\medskip
Together, these two crosscutting themes—AI to advance scientific productivity and sustainability, and community engagement with capacity building for next-generation scientific computing—represent both the engine and the scaffolding of future scientific computing. 
They are deeply interdependent: AI systems must be designed, deployed, and governed by engaged, well-supported communities; and those communities must be equipped with the tools, practices, and shared knowledge that AI can help accelerate. 
Advancing one without the other will limit impact. Investing in both will accelerate progress and ensure that next-generation ecosystems for scientific computing are not only more powerful, but also widely accessible, trusted, and sustainable.

\section{Required Research Directions and Community Actions}

Considering the challenges and opportunities outlined in Section~\ref{sec:challenges}, 
the workshop identified  research directions and community actions 
needed for accelerating progress in next-generation scientific computing. 
As discussed in Sections~\ref{sec:required-research-software} through \ref{sec:required-research-pedagogy}, 
each priority research area combines practical needs with aspirational goals, aiming to foster a productive interplay between human insight and AI capabilities while maintaining a strong foundation of scientific rigor and community engagement.
The recommended research directions emphasize key design principles: modularity, interoperability, trustworthiness, and extensive participation. 
The intent is to support both near-term implementation and long-term sustainability of emerging software and collaboration models for scientific computing. 
Section~\ref{sec:required-actions} outlines strategic community actions that are essential for supporting the software ecosystems, collaborative teams, and educational innovations discussed throughout this report. 
Taken together, the priority research directions and community actions define a comprehensive agenda for building thriving, resilient, and forward-looking scientific computing ecosystems.  

\subsection{Software ecosystems for AI in scientific computing}
\label{sec:required-research-software}

Next-generation scientific computing demands a radical rethinking of how software ecosystems are designed, developed, and sustained. 
As AI becomes more integrated into scientific workflows, software infrastructure must evolve to support new modes of discovery, while maintaining the rigorous standards of scientific correctness, reproducibility, and performance. 
This section identifies research priorities for building AI-integrated scientific software ecosystems that are modular, trustworthy, and adaptable. 
The topics below reflect a blend of foundational challenges and forward-looking opportunities, ranging from infrastructure and interoperability, to AI-assisted code generation, to robust validation methods that incorporate community knowledge and formal guarantees.
Taken together, these efforts can help create a robust, adaptive foundation for software ecosystems that meet the scale, complexity, and scientific integrity required for the AI-powered future of discovery.

\paragraph{Software and infrastructure for next-generation science.}  
Existing software infrastructure, although rich in abstractions, libraries, and tools, is inadequate for the multipronged challenges of next-generation scientific applications. 
The growing demands of high-fidelity models and the fragmentation of system software across AI-driven hardware platforms have created major obstacles for scientific teams. 
Libraries and tools may be underutilized because of the complexity of manually navigating multiple packages, and most science teams cannot feasibly manage the complexity of hybrid HPC/AI environments without new modes of support. Similar software challenges were articulated in the National Artificial Intelligence Research Resource (NAIRR) Software Workshop~\cite{nairr-software-workshop-report2025}.

A promising direction is to design software substrates that enable AI/ML and traditional simulation codes to interoperate scalably and adapt to evolving workflows, platforms, and algorithms. 
At present we have little understanding of what such  substrates might look like, but we can begin with
research 
to determine and minimize the intrusiveness required to connect legacy scientific software with new AI tools, coordinate multiple coexisting ecosystems (e.g., AI agents interfacing with scientific tools), and analyze how software components interact in real workflows. 
Of particular importance is exploring the use of AI agents in end-to-end scientific code development workflows, including interactions with math libraries, visualization tools, debuggers, job schedulers, and HPC performance analysis tools.   
Key areas of research include energy-efficient implementations of numerical and AI methods, tools for software usage analytics, AI-assisted debugging, 
and strategies for hybridizing traditional solvers with 
SciML methods.
Also needed are metrics to evaluate emerging SciML approaches against traditional numerical methods. 

\paragraph{Development of science-aware AI code generation systems.} A key
  research priority is the development of AI systems designed
  and trained specifically for scientific computing
  contexts. This requires more than adapting general-purpose LLMs; it calls for
  domain-specific training on high-quality scientific
  codebases, the incorporation of physics-informed constraints, and the use of validation datasets
  that emphasize correctness as well as functionality. Research should
  focus on developing hybrid approaches that combine traditional
  symbolic reasoning with neural methods, enabling AI systems to
  respect physical laws, dimensional analysis, and numerical stability
  requirements. 
Promising techniques include physics-informed neural networks (PINNs), neural operators, and differentiable programming, which offer pathways for AI-generated code to honor physical laws, conserve energy, and maintain numerical stability. Systems must be capable of constrained code generation (e.g., ensuring energy conservation in molecular dynamics) and must integrate domain-specific knowledge graphs that encode scientific principles.  
Formal verification methods tailored to scientific applications are also essential.
Key components include training datasets curated
for scientific applications with verified correctness and development of specialized architectures
that can reason about mathematical relationships and physical constraints. 
This research direction should also investigate methods for uncertainty quantification in AI-generated scientific code, enabling users to understand confidence levels and potential error propagation. 
We also need advances in AI interpretability, such as understanding AI decisions in safety-critical scientific applications.
To integrate existing code into new development, research is needed on AI tools that can
  generate, refactor, and modernize scientific code while ensuring   correctness and maintainability. Work is needed to use AI to extract knowledge from  existing codebases and generate useful, accurate documentation for  end users and contributors. AI evaluations of code bases can also be
  used to inform teams and leaders of risks in a code base (fragility,  limited testing, etc.) and to help teams make decisions about rewriting and  refactoring. We must also explore systems where human developers interactively train AI models in context, for example, by correcting
  recommendations.
  
\paragraph{Community-integrated validation and continuous learning frameworks.}
Given the critical importance of correctness in scientific software,
research is needed to develop 
validation frameworks that
leverage community knowledge, expert input, and formal guarantees. These systems should support automated generation of test cases grounded in scientific
principles, including testing against known analytical solutions where available.  
Techniques such as property-based testing and metamorphic testing are especially valuable for scientific domains, enabling validation of AI-generated code even in the absence of exact reference solutions. 
For example, fluid dynamics code produced by AI must satisfy divergence-free velocity fields and conserve mass properties that can be checked independently of a specific output. 
Research should also explore adaptive ``living" validation systems that evolve with
scientific understanding and incorporate domain expertise as part of the AI training feedback loop.
Peer review mechanisms tailored to code generation could facilitate expert vetting of outputs and structured incorporation of corrections into model refinement. 
These systems may include code that is validated against
observations and experiments that cannot be replicated.  
The use of formal methods for system correctness \cite{brooker2025SystemsCorrectness} and the
verification of implementations of numerical methods \cite{gorard2025shockconfidenceformalproofs}, as well
as AI methods 
\cite{chennabasappa2025llamafirewall,
nicolajsen2025ProgrammingInAgeOfAI,
schmied2025llmsgreedyagentseffects}, 
is needed.  
In addition, as AI-generated code and integrated workflows become more complex and widely adopted, research should address emerging security concerns, including software vulnerabilities and ecosystem-level threats. 
These approaches should complement continuous integration and testing practices, forming a robust set of validation methods that can sustain accuracy, trust, and progress as AI becomes increasingly integrated into scientific computing.

\subsection{Cross-disciplinary collaboration and AI for scientific software teams}
\label{sec:required-research-teams}

The challenges of next-generation scientific computing demand teams that can operate across disciplinary boundaries, institutional cultures, and evolving technological landscapes. 
AI adds a new dimension to this complexity, both as a source of capability and as a potential collaborator. 
Research in this area must explore how scientific teams can effectively integrate AI systems into their workflows, while maintaining human creativity, trust, and scientific rigor. This includes developing new frameworks for trust, formalizing human-AI team roles, and applying lessons from prior collaborations in science, social science, and the private sector to guide effective engagement throughout the software lifecycle.

\paragraph{A multilevel model of trust: Teams and technology.} 
Trust—foundational to effective collaboration—has been studied extensively as a concept; rich literature exists describing its presence in teams, differences between trust and distrust, and its application to technologies including those endowed with autonomy \cite{Feitosa2020, Dirks2022, Hancock2023}. 
A multilevel, dynamic model of trust is needed to guide cross-disciplinary teams working in hybrid HPC/AI environments.  
This includes constructs such as propensity to trust, perceived trustworthiness, 
trust in AI, trust in automation and automated systems, transparency, and reliance.
Such a model must account for interactions across individuals, teams, and technologies, helping uncover opportunities to build and maintain trust, while also reducing the likelihood of cascading failures caused by misplaced trust or unaddressed uncertainty.

\paragraph{Roles for AI: Teammate, trainer, tool.}
As AI becomes increasingly embedded in scientific research, clarifying its roles within and across 
teams becomes essential. 
We must examine modes of human-AI teaming that augment human creativity and decision-making, especially in scientific code development and debugging. 
For example, AI-assisted collaboration tools could assist with 
translation between domain vocabularies (such as between scientific notation used by physicists and programming constructs familiar to research software engineers).
To effectively integrate AI in teams and mitigate the likelihood of negative outcomes associated with the adoption of AI, research is needed to formalize and classify the various forms of taskwork and teamwork \cite{nas-teams2025} in scientific computing. 
Doing so could enable, for example, the development of an adaptive taxonomy of roles and tasks for AI that is informed by the capabilities and limitations of the technology, the needs of the team, and the particulars of the domain. 
This configurable knowledge base could guide team design, risk management, and human-AI interaction strategies across the software  lifecycle.

\paragraph{Effective engagement within cross-disciplinary teams.} 
We must draw on insights from past scientific computing projects and other endeavors to foster environments that support extensive participation and cross-disciplinary success in next-generation computational science~\cite{Fiore2008, KoehlerLeman2020, ideas-ecp2024}.
This includes updating institutional policies and incentives to recognize collaborative work~\cite{Howison2013} and embedding communication norms~\cite{Milewicz2018}, governance structures, and knowledge-sharing practices that welcome contributions from a wide range of stakeholders throughout the entire software lifecycle—including funding acquisition; software design, development, and testing; developer and user (re)training; and use and maintenance of the software itself to conduct science. 
Efforts to codify best practices, define success metrics, and incentivize adoption are essential for building resilient, high-performing teams for the AI-augmented scientific future.

\subsection{Pedagogy and workforce development in the age of AI}
\label{sec:required-research-pedagogy}
 
A perennial shortage of well-qualified scientists and engineers with the breadth of knowledge needed for today's and tomorrow's scientific computing underscores the urgent need for new pedagogical approaches, training programs, and workforce strategies. 
These efforts must support the development of cross-disciplinary skills, 
incentivize and promote ways to integrate different methodologies, 
ease onboarding into complex research projects, and enable continuous learning to keep pace with rapid technological evolution.

\paragraph{Workforce development and human-AI collaboration models.}
  Research is urgently needed to understand how workforce
  development for scientific computing must evolve in AI-augmented environments. This
  includes empirical studies of how AI tools influence learning, skill
  development, and scientific reasoning among students and
  researchers at different career stages. Key questions
  include the following: Which fundamental skills remain essential when
  AI handles routine coding tasks? How can pedagogical approaches
  combine AI assistance with deep understanding of
  scientific principles?  How should assessment methods evolve to
  evaluate comprehension in AI-assisted contexts? This research
  should also explore new models of human-AI
  collaboration in scientific software development, including optimal
  division of labor, interface design for
  effective teaming, and strategies for maintaining human expertise
  and critical thinking while leveraging AI's
  computational advantages. 
  
\paragraph{Cross-disciplinary curriculum design.} Educational
programs must integrate AI, software engineering, and scientific
domain knowledge—from K-12 through graduate training.
Identifying key competencies (e.g., trustworthy and fair AI,
reproducibility, teamwork) is essential for preparing future software developers in scientific fields. 
Curriculum development must become more flexible  and modular, enabling faster iteration and responsiveness to the changing landscape. 
This includes fostering partnerships across academia, national labs, and the private sector. 
Professional development programs should also address cross-disciplinary collaboration skills and 
provide transformational hands-on experiences that illustrate the impact and excitement of teamwork in scientific computing, 
while fostering collaborative  environments that respect and adapt to different domain cultures.

\paragraph{Innovative curricula and programs incorporating socio-technical co-design.} New instructional models are needed that emphasize
project-based learning, especially through collaborations among the private sector, academia, and national laboratories. 
These models should extend beyond STEM domains to include the humanities and other relevant disciplines, ensuring a broader understanding of scientific and societal impact.  The models should also include thoughtful integration of career pathways to support entry at multiple points, meeting people where they are, rather than requiring people to enter with a specific set of pre-defined prerequisites.
In addition, we should capture lessons from successful efforts.  For example, the collaboration model adopted by the Exascale Computing Project, in partnership with DOE leadership computing facilities, was highly successful in launching the exascale era by co-creating scientific applications, human capital, and a robust scientific software ecosystem. Not only did ECP bridge research and computing facilities in a co-design approach, ECP included early attempts at socio-technical co-design when it added talent development through the Sustainable Research Pathways~\cite{shi-website} program and training pipelines through the  Center for Scientific Collaboration and Community Engagement (CSCCE)~\cite{cscce-website}.
  
\paragraph{Community
  and workforce success metrics.}  
Clear, ambitious metrics are needed to measure progress in building the HPC+AI workforce. 
For example, a pilot program to advance the workforce for next-generation scientific computing could aim to increase the talent pool by at
least 1,000 people over five years, averaging more than 200 new and upskilled contributors annually who are equipped to tackle emerging challenges in scientific software ecosystems.
Well-defined metrics can help align national efforts, guide strategic investments, and ensure that workforce development keeps pace with evolving technical demands.

\subsection{Required community actions}
\label{sec:required-actions}

While technical research is a key driver of innovation in scientific computing, the workshop emphasized that progress in foundational scientific software also relies on sustained collaboration across institutions, disciplines, and sectors. Building trustworthy, capable, and sustainable software ecosystems requires more than technical breakthroughs—it calls for strategic coordination, shared infrastructure, supportive policies, and workforce investment.
Various organizations, projects, and 
foundations—such as 
the Consortium for the Advancement of Scientific Software, 
High Performance Software Foundation, 
IDEAS project, 
NumFOCUS foundation, 
Research Software Alliance, 
Software Sustainability Institute, and
US Research Software Engineering Organization—are
making meaningful strides in shifting the cultural and structural foundations of scientific software~\cite{community-organizations:2019}.  
Raising awareness of these efforts and contributing to them—whether directly or through complementary action—will benefit the broader scientific community.  Much work remains to be done, however, particularly in advancing high-impact partnerships and addressing persistent gaps in infrastructure, recognition, and career paths.
 
\paragraph{Strategic public-private partnerships.} 
Of the many strategies discussed, this emerged from the workshop as a central priority. 
Stronger collaboration among academic institutions, national laboratories, and the private sector is essential to the future of AI-enabled scientific software. 
These partnerships must go beyond short-term research goals to support co-designed solutions that align with long-term scientific needs and technological shifts.  
The private sector includes both commercial developers of computing technologies—such as hardware manufacturers, cloud service providers, and AI platform companies—and major industrial users of scientific software. 
Companies across the automotive, aerospace, energy, manufacturing, and materials sectors (e.g., Ford, GE Research, Boeing, and others) depend on high-performance computing and simulation to support product design, optimization, and predictive modeling. Their engagement brings valuable application-driven insights, real-world performance requirements, and opportunities to accelerate the translation of foundational advances in computing into practical tools.
Sustained coordination is needed to shape not only the technical requirements of software tools but also the practices and training ecosystems that support their adoption.
Joint efforts might include cross-sector learning communities, affinity groups, shared training initiatives, and advisory partnerships that bring together HPC practitioners, AI developers, and domain scientists.
Of particular importance is the need to align the rapidly evolving commercial AI and hardware landscapes with the long-term needs of the scientific community. 
These shared efforts play a vital role in maintaining national competitiveness, advancing research, and ensuring that science, engineering, and manufacturing progress together for long-term socioeconomic resilience.

\paragraph{Community engagement and broad-access infrastructure.} 
As scientific software projects grow in scale and complexity, they require engagement models that foster trust, transparency, and shared ownership across teams and institutions. 
Building strong, multi-institutional collaborations depends on clear frameworks that support distributed leadership and far-reaching participation. 
Key strategies include implementing contributor agreements, project charters, and open governance models that reflect a wide range of stakeholder voices. Equally critical is sustained effort to ensure extensive, fair access to infrastructure and participation. This means funding and scaling infrastructure projects that lower barriers for under-resourced institutions and communities—such as through low-cost cloud access, peer mentoring networks, and national initiatives.

\paragraph{Recognition, incentives, and long-term community investment.} 
The sustainability of scientific software ecosystems depends on valuing and supporting the full range of essential contributions, including development, documentation, user support, training, maintenance, and community stewardship. These roles must be reflected in hiring, promotion, and funding decisions and  reinforced through long-term investment strategies. Stable, visible support for these roles strengthens the resilience of the overall ecosystem and builds the capacity needed for lasting progress. 

\paragraph{Responsible innovation guidelines and fair AI.} 
Advancing AI-driven scientific discovery without oversight regarding fairness risks opening a Pandora’s box of unintended consequences. Community-wide guidelines for the responsible integration of AI are essential and must be treated as a top priority. These guidelines should address algorithmic transparency, data governance, and reproducibility, while also establishing norms to prevent bias, protect users, and anticipate unintended consequences. 
A key part of this effort is ensuring that datasets are complete and representative of the full range of individuals and perspectives they aim to serve.
Responsible innovation must remain a core design principle as both tools and teams evolve.

\bigskip
\noindent
These policy and collaborative efforts are not secondary to technological innovation—they are foundational. 
By prioritizing strategic partnerships and reinforcing them with community-driven practices, institutional reforms, and long-term investment, 
while considering the long-term needs of both the nation’s research and industrial sectors,
we can 
address the evolving demands of software- and AI-driven science.
Ultimately, this integrated approach will foster 
ecosystems where technological breakthroughs and social progress advance hand in hand, reinforcing one another 
for a more innovative, responsible, and resilient future.

\section{Conclusion and Next Steps}

Scientific computing is entering a new era—driven by the integration
of AI, the complexity of cross-disciplinary collaboration, and the
growing  demand for scalable, sustainable software ecosystems.
As introduced in Figure~\ref{fig:socio-technical-co-design}, advancing
next-generation scientific computing requires a socio-technical
co-design approach to fully harness community, software, and AI for
cross-disciplinary team science.  This means not only developing new
software and hardware but also creating systems of
practice—collaboration norms, training models, and governance
strategies—that evolve with the technology.  We launched a series of
workshops to begin a dialogue in the computational science community
and other related disciplines on how to confront and resolve these
challenges. This report summarizes the outcomes of the first  
workshop in the series, representing the first step in a
multiyear effort to understand how to co-design next-generation
ecosystems for scientific computing. 
We envision ecosystems that are not only technologically advanced but
also grounded in shared responsibility, fair practices, and a
renewed commitment to bold, visionary science.

The workshop identified challenges and opportunities across three
core domains: software infrastructure, cross-disciplinary
collaboration, and pedagogy.  Two
crosscutting themes emerged: the transformative role of AI and the
foundational importance of workforce and community development. 
Addressing these requires coordinated action on several fronts: 
advancing modular AI-integrated software; developing practices and tools that foster effective cross-disciplinary collaboration; and reimagining training for hybrid human-AI environments. 
Equally critical is aligning institutional policies, governance models, and community norms to ensure the long-term sustainability and growth of scientific software ecosystems and communities.
Pilot projects can catalyze progress in these areas, with priorities evolving over time:
\begin{itemize} 
\vspace{-0.06in}
\item {\bf Near-term (1–2 years)}: Launch pilot programs on hybrid AI/HPC software infrastructure as well as cross-disciplinary
collaboration and pedagogy for AI-driven scientific computing, establish responsible AI guidelines, and prototype public–private partnerships.
\vspace{-0.06in}
\item {\bf Mid-term (3–5 years)}: Explore scaling approaches for modular AI/HPC software ecosystems, launch workforce training curricula, expand community development, implement and evaluate several institutional policy reforms.
\vspace{-0.06in}
\item {\bf Long-term (5+ years)}: Develop and evaluate globally networked ecosystems for AI-driven scientific computing, explore frameworks for community governance at scale, seek ways to integrate AI agents as active research collaborators.
\end{itemize}

Evolving next-generation scientific computing
ecosystems, with many technical and social
requirements, may seem daunting. However, the challenges also bring
opportunities to learn from past successes and plan for agile
approaches to 
grow communities and workforces poised to
evolve alongside the changing landscape of HPC technologies and AI.
Times of great change are marked by both enormous promise and significant uncertainty. 
Meeting this moment will require not just incremental improvements but bold, risk-tolerant thinking. 
Just as the dawn of quantum physics revolutionized our understanding
of the universe, today’s scientific landscape demands a return to that
spirit of curiosity, invention, and courage to chart new waters. 
Human ingenuity—coupled with emerging AI capabilities—must be placed at the center of scientific progress.
By intentionally interweaving technical and social components in next-generation scientific computing, we can create feedback loops that accelerate both scientific discovery and real-world impact.

\phantomsection
\section*{Acknowledgments}
\addcontentsline{toc}{section}{Acknowledgments}

This workshop was partially
supported by the U.S. Department of Energy (DOE) Office of Science Distinguished Scientist Fellows Program.  We especially thank our DOE contacts: Hal Finkel and David Rabson, DOE Office of Advanced Scientific Computing Research (ASCR).
  
We thank Katie Antypas (National Science Foundation, Office of Advanced Cyberinfrastructure), April Hanks (Team Libra) and Christina Mihaly Messina
(SNL) for insightful contributions to workshop discussions on next-generation ecosystems in scientific computing, which helped shape the ideas conveyed in this report.
We are grateful to Suzanne Parete-Koon (ORNL) for detailed feedback on the document; her suggestions improved the precision and perspective of the report. 
We thank Gail Piper for editing this manuscript.

\newpage
\clearpage
\phantomsection
\addcontentsline{toc}{section}{References}

\bigskip
\bigskip

\appendix

\newpage
\section{Glossary}

This report aims to communicate across a broad community, including experts in high-performance computing, AI, compuational science, social and cognitive science, team science, community and workforce development, and related topics. To support our goal of communicating clearly across disciplines, this glossary defines terms as used in this report.  

\begin{itemize}
\item {\bf AI for science}: The next generation of methods and scientific opportunities in computing, including the development and application of AI methods (e.g., machine learning, deep learning, statistical methods, data analytics, automated control, and related areas) to build models from data and to use these models alone or in conjunction with simulation and scalable computing to advance scientific research~\cite{AI-for-science2020}.

\item {\bf Cross-disciplinary}: Collaboration among multiple disciplines toward a shared objective~\cite{nas-teams2025}.  

\item {\bf Ecosystems for scientific computing}: Dynamic socio-technical systems made up of people, technologies, infrastructure, institutions, workflows, and cultural practices that collectively support the development and evolution of scientific computing. [page~\pageref{def:ecosystems} of this report]:

\item {\bf Science of team science}:  Research area in which scholars from various disciplines, such as psychology, organizational sciences, sociology, communication, and philosophy, contribute conceptually and empirically to understanding how science teams are organized and work together, how to best measure their effectiveness, and the implications of individual differences in team science~\cite{nas-teams2025}.

\item {\bf Socio-technical co-design for scientific computing}:  The intentional and integrated development of both technical components (e.g., software, AI, infrastructure) and social components (e.g., teams, institutions, practices, training) of ecosystems for scientific computing, ensuring that social and technical dimensions are addressed not in isolation but as tightly coupled elements of a unified, forward-looking strategy~[page~\pageref{sec:socio-technical-co-design} of this report].

\item {\bf Taskwork}: Activities associated with achieving a team’s goals~\cite{nas-teams2025}.

\item {\bf Teamwork}: Interactions among team members that are essential for effective collaboration~\cite{nas-teams2025}.

\item {\bf Team science}: Collaborative, interdependent research conducted by more than one individual~\cite{nas-teams2025}.

\end{itemize}

\medskip
\noindent
{\bf Acronyms}

\begin{itemize}
\item {\bf AI}: Artificial intelligence
\item {\bf HPC}: high-performance computing
\item {\bf CPU}: Central processing unit
\item {\bf GPU}: graphics processing unit
\item {\bf STEM}: science, technology, engineering, and mathematics
\item {\bf NSF}: National Science Foundation
\item {\bf DOE}: U.S. Department of Energy
\end{itemize}

\newpage
\section{Workshop Description}
\label{sec:workshop-description}

The 2025 Workshop on {\em Next-Generation Ecosystems for Scientific Computing: Harnessing Community, Software, and AI for Cross-Disciplinary Team Science} engaged over 40 cross-disciplinary experts in Chicago, IL from April 29 to May 1, 2025, to chart a path toward more powerful, sustainable, and collaborative scientific software ecosystems.

\subsection*{Workshop Charge}

\begin{wrapfigure}{R}{0.37\textwidth}
\vspace{-0.4in}
\includegraphics[width=0.37\textwidth]{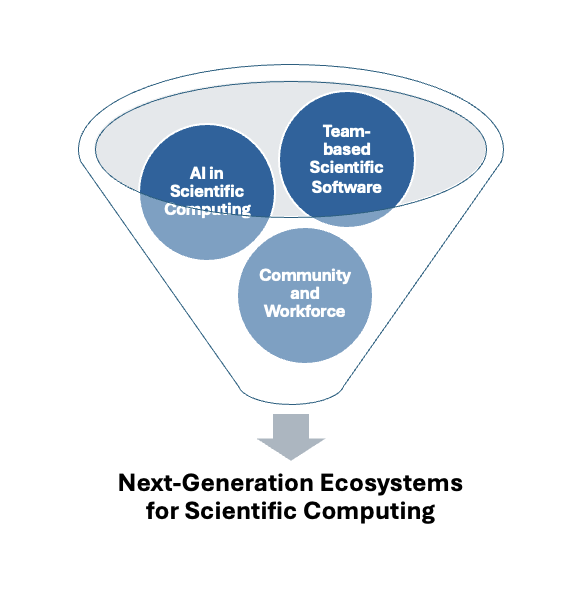}
\vspace{-0.4in}
\end{wrapfigure}
The HPC community has long been a leader in advancing scientific discovery to new frontiers. High-quality scientific software—which encapsulates expertise across disciplines for use by others—is a primary means of sustained collaboration and scientific progress.  Motivated by urgent issues raised in recent community reports on the state of scientific software development and AI for science, energy, and security, the workshop will bring together cross-disciplinary experts—in HPC, AI, computational science, applied math, computer science, research software engineering, cognitive and social sciences, and community development—to identify challenges, prioritize gaps, and explore opportunities that will shape next-generation ecosystems for scientific computing. The workshop will follow a co-design methodology that intentionally weaves together topics in team-based scientific software, AI in scientific computing, and community/workforce development, thereby ensuring a holistic approach. 

\paragraph{Areas of emphasis.} The workshop's goal is to assess and transform team-based scientific software, with emphasis on building ecosystems to address the needs of next-generation research in scientific computing, while advancing emerging AI technologies.  We will integrate technical and community considerations throughout discussions on:

\begin{titemize}
\item {\bf Software and next-generation science}
\begin{titemize}
\item Exploring how new scientific frontiers and heterogeneous computing architectures require new approaches for software and workforce development.
\item Incorporating community-building strategies to ensure that new technical solutions reflect the perspectives and expertise of all stakeholders representing a broad swath of backgrounds.
\end{titemize}

\item {\bf Software ecosystems for AI in scientific computing}

\begin{titemize}
\item Mapping pathways to create robust, scalable software ecosystems that catalyze AI-driven discoveries in HPC contexts and produce increased innovation.
\item Developing more robust AI tools and resources for scientific computing that address current gaps and include wide-ranging perspectives.
\end{titemize}

\item {\bf Team-based software and cross-disciplinary research}

\begin{titemize}
\item Identifying roles, methodologies, and best practices for team-based scientific software.
\item Emphasizing community co-design, where software solutions evolve through iterative dialogue among scientists, developers, and end users.
\end{titemize}

\item {\bf AI for scientific software productivity and sustainability}

\begin{titemize}
\item Addressing how varied perspectives and AI can accelerate development, use, performance, and impact of scientific software.
\item Embedding community-based feedback loops to align AI-driven tools with real-world needs and foster broader engagement.
\end{titemize}

\item {\bf Community and workforce development}

\begin{titemize}
\item Accelerating strategies to cultivate future-generation R\&D teams in the computing sciences.
\item Fostering collaborative environments that tap into the strengths of a broad workforce, bridging technical and community-building efforts.
\end{titemize}
\end{titemize}

\subsection*{Workshop Objectives} 

We aim to shape a forward-looking plan for the future of team-based software in scientific computing, along with actionable strategies to bring it to life, all while cultivating a vibrant community.  Key areas of focus—considering emerging AI technologies and the needs of cross-disciplinary research—will include:

\begin{titemize}

\item
{\bf Understanding scientific software practices}: Building a shared understanding of methodologies for team-based software in HPC, explicitly integrating technical and community perspectives.

\item {\bf Characterizing roles and success factors}: Identifying critical roles in scientific software teams and community-based success factors that sustain effective software development and user engagement.

\item {\bf Identifying challenges and opportunities}: Examining gaps (both technical and community-driven) that hinder team effectiveness and highlighting opportunities to address emerging research needs.

\item {\bf Envisioning the future}: Considering bold ideas for next-generation scientific software, emphasizing synergy between technical solutions and community dynamics.

\item {\bf Fostering community development}: Developing actionable steps to strengthen the workforce and build a vibrant software community prepared to meet urgent challenges in HPC and AI-driven scientific computing.

\end{titemize}

\noindent
By intentionally blending technical and community factors, the workshop aims to shape a future where thriving, cross-disciplinary communities drive the next wave of scientific discovery, with high-quality scientific software as a keystone of sustained collaboration and scientific progress.

\subsection*{Workshop Outcomes}

Participants will map the landscape of team-based scientific software, characterize key roles and success factors, prioritize focus areas, and develop strategies to shape the future of scientific computing. These integrated insights—to be documented in a post-workshop report—will inform future efforts and may drive changes in perspectives, policies, and practices throughout the scientific computing community.

This workshop marks the first of a three-year series dedicated to strengthening team-based scientific software in an AI-driven future, with each year incorporating the co-design framework to ensure that both technical and community needs are addressed:

\begin{titemize}
\item {\bf Year 1 emphasis}: Identifying and understanding challenges, gaps, and opportunities
\item {\bf Year 2 emphasis}: Developing and evaluating strategies and progress
\item {\bf Year 3 emphasis}: Coordinating ecosystem-wide advancements that meld technical solutions and community-building
\end{titemize}

\noindent
By intentionally weaving community considerations into technical discussions—and vice versa—this three-year series will continuously expand the community base and help to advance next-generation scientific discovery.

\newpage
\section{Workshop Participants}
\label{sec:workshop-participants}

\subsection*{Organizing Committee}

\begin{titemize}

\item Lois Curfman McInnes, Argonne National Laboratory (ANL), Senior Computational Scientist,
Mathematics and Computer Science (MCS) Division

\item Dorian Arnold, Emory University, Associate Professor, Department Computer Science

\item Prasanna Balaprakash, Oak Ridge National Laboratory (ORNL), Director of AI Programs 

\item Mike Bernhardt, Team Libra, Founder and Chief Strategy Officer; 
former Director of Communication for the DOE Exascale Computing Project

\item Beth Cerny, ANL, Head of Communications for the Argonne Leadership Computing Facility (ALCF)

\item Anshu Dubey, ANL, Senior Computational Scientist, MCS Division

\item Denice Ward Hood, University of Illinois Urbana-Champaign, Associate Professor, Department of
Education Policy, Organization \& Leadership

\item Mary Ann Leung, Sustainable Horizons Institute, Founder and President

\item Olivia B. Newton, University of Montana, Faculty, Department of Management Information Systems

\item Stefan M.\ Wild, Lawrence Berkeley National Laboratory (LBNL), Director, Applied Mathematics and Computational Research Division
\end{titemize}

\subsection*{Attendees}

\begin{titemize}
\item Katie Antypas, National Science Foundation, Director of the Office of Advanced Cyberinfrastructure

\item Tony Baylis, Lawrence Livermore National Laboratory (LLNL), Senior Manager

\item  David Bernholdt, ORNL, Distinguished R\&D Staff Member, Computer Science and Mathematics (CSM)
Division

\item Chris Cama\~{n}o, California Institute of Technology, Ph.D. Student, Applied and Computational Mathematics

\item Hannah Cohoon, LBNL, User Experience Researcher, Scientific Data Division
 
\item Charles Ferenbaugh, Los Alamos National Laboratory (LANL), R\&D Scientist, Applied Computer Science

\item Hal Finkel, DOE Office of Advanced Scientific Computing Research, Director of Computational Science
Research and Partnerships
 
\item Stephen M. Fiore, University of Central Florida, Professor, Director of Cognitive Sciences Laboratory,

\item Sandra Gesing, US Research Software Engineers Association, Executive Director; Senior Researcher,
San Diego Supercomputer Center
 
\item Roscoe Giles, Boston University, Professor, Electrical and Computer Engineering, also Computing and Data Science

\item Diego G\'{o}mez-Zar\'{a}, University of Notre Dame, Assistant Professor, Computer Science \& Engineering

\item April Hanks, Team Libra, Director of Event Services

\item James Howison, University of Texas at Austin, Associate Professor, School of Information
 
\item Tanzima Islam, Texas State University, Assistant Professor, Department of Computer Science

\item David Kepczynski, Ford Motor Company, Global Chief Digital Transformation

\item Charles Lively, LBNL, Science Engagement Engineer and HPC Consultant, NERSC
 
\item Vanessa López-Marrero, Brookhaven National Laboratory / Stonybrook University, Computational Scientist

\item Harshitha Menon, LLNL, Computer Scientist, Center for Applied Scientific Computing,

\item Bronson Messer, ORNL, Director of Science for OLCF

\item Christina Mihaly Messina, SNL, HPC Developer \& Metadata Manager, ORS Program

\item Paul Messina, ANL, Argonne Associate and Distinguished Fellow, Computing, Environment and Life Sciences

\item Marieme Ngom, ANL, Assistant Computer Scientist, ALCF

\item Christopher Oehmen, Pacific Northwest National Laboratory, Senior Research Scientist and Group
Leader, Biological Sciences Division

\item Umesh Paliath, GE Aerospace Research, Technology Manager for Aerodynamics and CFD Methods

\item Michael Papka, ANL, Deputy Associate Laboratory Director, Computing, Environment and Life Sciences;
Director of ALCF; University of Illinois Chicago, Professor, Computer Science

\item Irene Qualters, LANL, Associate Laboratory Director Emerita, Simulation and Computation

\item David Rabson, Department of Energy, Advanced Scientific Computing Research, Physical Scientist

\item Elaine M. Raybourn, Sandia National Laboratories (SNL), Social Scientist, Engineering \& Software
Research

\item Katherine Riley, ANL, Director of Science for ALCF

\item Paulina Rodriguez, The George Washington University, Ph.D. Student, Mechanical and Aerospace
Engineering

\item Damian Rouson, LBNL, Group Lead for Computer Languages and System Software, AMCR Division

\item Michelle Schwalbe, National Academies of Sciences, Senior Director, Board on Mathematical Sciences
and Analytics, Board on National Materials and Manufacturing

\item Sudip Seal, ORNL, Group Lead for Systems and Decision Sciences, CSM Division

\item \"{O}zge S\"{u}rer, Miami University, Assistant Professor, Department of Information Systems and Analytics

\item Valerie Taylor, ANL, Director of the Mathematics and Computer Science (MCS) Division 

\item Jim Willenbring, SNL, Senior Member of R\&D Technical Staff, Center for Computing Research

\item Lou Woodley, Center for Scientific Collaboration and Community Engagement, Founder and Director

\item Lingfei Wu, University of Pittsburgh, Assistant Professor, Department of Informatics and Networked
Systems

\end{titemize}

\section{Workshop Agenda}
\label{sec:workshop-agenda}

\includegraphics[scale=0.82,page=1,trim={0.8cm 0cm 2.0cm 1.5cm},clip]{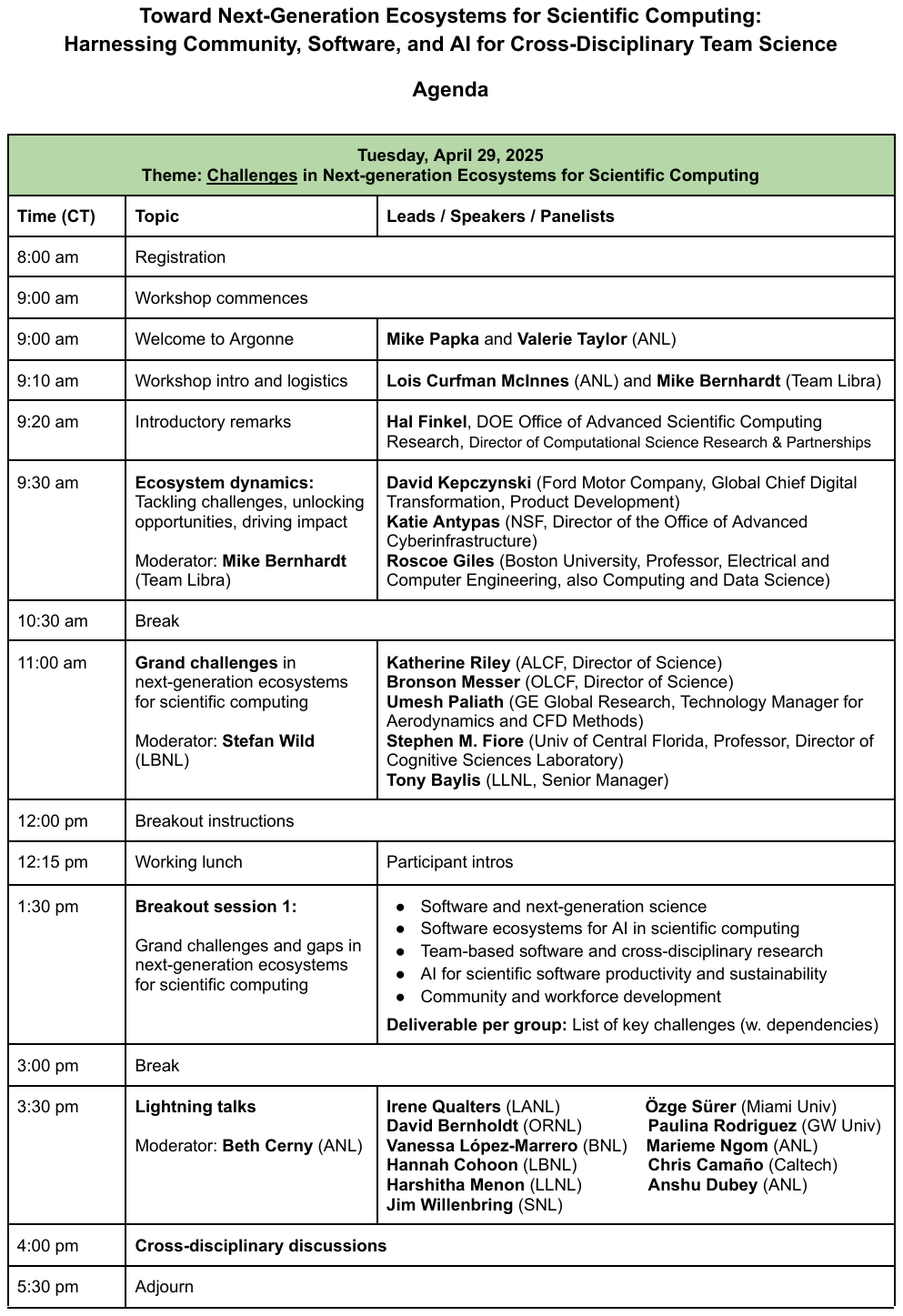}
\newpage
\includegraphics[scale=0.82,page=2,trim={1.4cm 0cm 2.0cm 1.5cm},clip]{figures/Agenda.Workshop.Next-GenEcosystems.2025.04.pdf}
\newpage
\includegraphics[scale=0.82,page=3,trim={1.4cm 0cm 2.0cm 1.5cm},clip]{figures/Agenda.Workshop.Next-GenEcosystems.2025.04.pdf}.

\end{document}